%% file: paper.tex
\documentclass[10pt]{article}
\usepackage{amsmath,amssymb}
\newcommand{\ls}{{\it l}} 
\newcommand{\ZZ}{{\mathbb Z}} 
\newcommand{\QQ}{{\mathbb Q}} 
\newcommand{\divides}{\mid}
\newcommand{\notdivides}{\nmid}
\makeatletter
\def\GCD{\qopname\relax\@empty{GCD}} 
\def\lc{\qopname\relax o{lc}} 
\makeatother

\newtheorem{theorem}{Theorem}
\newtheorem{definition}{Definition}

\newtheorem{lemma}{Lemma}

\newcommand{\Q}{\QQ}
\newcommand{\sQ}{\QQ}
\newcommand{\Z}{\ZZ}

\newcommand{\Fp}{{\mathbb F}_p}
\newcommand{\degr}{d}
\newcommand{\x}{x}
\newcommand{\minp}{m}
\newcommand{\bad}{lc-bad }
\newcommand{\bbad}{lc-bad}
\newcommand{\EuclAlg}{{\rm GCD}}

\author{
  Michael Monagan\thanks{Supported by NSERC of Canada
  and the MITACS NCE of Canada.}\footnotemark[1] \quad
  Mark van Hoeij\thanks{Supported by NSF grant
  0098034.}\footnotemark[2] \\[10pt]
 \footnotemark[1]
  Department of Mathematics, Simon Fraser University, \\
  Burnaby, B.C., V5A 1S6, Canada. \\[10pt]
 \footnotemark[2]
  Department of Mathematics, Florida State University, \\
  Tallahassee, FL 32306-4510, USA.
}

\title{A Modular Algorithm for Computing
Polynomial GCDs over Number Fields presented with Multiple Extensions.}
\date{November 2005}

\begin{document}
\maketitle
\begin{abstract}
We consider the problem of computing the monic gcd of two
polynomials over a number field $L = \QQ(\alpha_1,\ldots,\alpha_n)$.
Langemyr and McCallum have already shown how Brown's modular
GCD algorithm for polynomials over $\QQ$ can be modified to work
for $\QQ(\alpha)$ and subsequently, Langemyr extended the
algorithm to $L[x]$.  Encarnacion also showed how to use
rational number to make the algorithm for $\QQ(\alpha)$ output sensitive,
that is, the number of primes used depends on the size of the integers
in the gcd and not on bounds based on the input polynomials.

Our first contribution is an extension of Encarnacion's modular GCD
algorithm to the case $n>1,$ which, like Encarnacion's algorithm,
is is output sensitive.

Our second contribution is a proof that it is not necessary
to test if $p$ divides the discriminant.
This simplifies the algorithm; it is correct without this test.

Our third contribution is a modification to the algorithm to treat
the case of reducible extensions.  Such cases arise when
solving systems of polynomial equations.


Our fourth contribution is an implementation of the
modular GCD algorithm in Maple and in Magma.
Both implementations use a recursive dense polynomial data structure for
representing polynomials over number fields with multiple field
extensions.

Our fifth contribution is a primitive fraction-free algorithm.
This is the best non-modular approach.
We present timing comparisons of the Maple and Magma implementations 
demonstrating various optimations and comparing them with the monic Euclidan
algorithm and our primitive fraction-free algorithm.

\end{abstract}

\section{Introduction}
We recall the relevant details of the so called {\em modular GCD algorithm}
first developed by Brown in \cite{Brown} for polynomials over $\ZZ$ and then
by Langemyr and McCallum in \cite{Langemyr}, Langemyr in \cite{AAECC} and
Encarnacion in \cite{Encarnacion} for polynomials over $L = \QQ(\alpha)$,
which we shall generalize to $L = \QQ(\alpha_1,\ldots,\alpha_n)$.
First some notation.

We denote the input polynomials by $f_1$ and $f_2$, their monic gcd by $g$.
The {\em cofactors} are the polynomials $f_1/g$ and $f_2/g$.
The {\em denominator} ${\rm den}(f)$ of $f \in \QQ[x]$ is the
smallest positive integer such that ${\rm den}(f) \, f \in \ZZ[x]$.
See section~\ref{sectionEucl} for the definition of ${\rm den}(f)$
if $f \in \QQ(\alpha_1,\ldots,\alpha_n)[x]$.
The {\em height} $H(f)$ of $f$ is the magnitude of the largest
integer appearing in the rational coefficients of $f$.

The {\em associate} $\tilde f$ of $f$ is defined as
$\tilde{f} = {\rm den}(h)\,h$ where $h = {\rm monic}(f)$.
Here ${\rm monic}(f)$ is defined as ${\rm lc}(f)^{-1} f$ where ${\rm lc}(f)$ is
the {\em leading coefficient} of $f$.
Define the {\em semi-associate} $\check{f}$ as $r \, f$ where
$r$ is the smallest positive rational for which ${\rm den}(r \, f)=1.$

\bigskip
\noindent
{\bf Examples:} If $f = 2 x - 4/3$ then
${\rm den}(f) = 3$, $H(f) = 4$, and $\check{f} = \tilde f = 3x - 2.$
If $\alpha= \sqrt 2$ and $f = - \alpha x + 1$ then $H(f) = 1$, $\check{f} = f$,
${\rm monic}(f) = x - \alpha/2$ and $\tilde f = 2 x - \alpha$.

\bigskip
\noindent
Computing the associate $\tilde{f}$ is useful for removing denominators,
but could be expensive if ${\rm lc}(f)$ is a complicated algebraic number.
So we preprocess the input polynomials in our algorithm
by taking the semi-associate instead.
If ${\rm lc}(f) \in \QQ$ then the two notions are the same up to a sign:
\[ \check{f} = \pm \tilde{f} \Longleftrightarrow {\rm lc}(f) \in \QQ \]

\subsection{Motivation for the algorithm}
\label{advantages}
The goal of this paper is to present an efficient
GCD algorithm over a field $L$ that consists of
multiple extensions over $\QQ$ that is practical.
As a motivating application, consider the problem of factoring $f \in L[x]$
using Trager's algorithm \cite{Trager}.
One sequence of gcd computations in $L[x]$ is required to compute the
square-free factorization of $f,$ beginning with ${\rm gcd}(f,f')$.
Then for each square-free factor, a second sequence of gcd computations
in $L[x]$ occurs when the irreducible factors of $f$ are determined.

Let $L$ be a number field of degree $D$ over $\QQ$ and
let $f_1, f_2 \in L[x]$ both have degree $n$ and let $g$ be their monic gcd.
For a computer algebra system to be effective at performing computations
in $L[x]$ we require a GCD algorithm for computing $g$ with a complexity
which is comparable to that of multiplication and division in $L[x].$
It is well known that the size of the integers in the
coefficients of the remainders in the Euclidean algorithm grows
rapidly and consequently, the Euclidean algorithm becomes ineffective
when $\deg g$ is much smaller than $n,$ the worst case being when $g = 1$.
This leads us to consider a modular GCD algorithm.

Let $c = H(g),$ that is, $c$ is is the magnitude of the largest
integer coefficient appearing the rational coefficients of $g.$
If we knew $c$ in advance, we could choose a single prime $p > 2 c^2$ from
a table, compute one modular image in $O(n^2 D^2 \log^2 p))$ time
and reconstruct the rational coefficients of $g$ in $O( n D \log^2 p )$ time.
However we do not know $c$ and accurate bounds are not possible when $c$
is much smaller than $H(f_1)$ and $H(f_2)$.
Thus we compute $g$ modulo a sequence of primes of almost constant bit length
and incrementally reconstruct $g$.
If we want ${\rm log}(m)=O({\rm log}(c))$, that is, if we want
the number of primes used to be proportional to the size of
the coefficients in $g$ so that small gcds are recovered quickly,
then we are forced to
\begin{enumerate}

\item Not use a primitive element to convert to a single extension,
which is expensive and can cause a blowup in the size of the coefficients.
This problem is well known, e.g. see \cite{Abbott}.  Note, although the conversion
to a primitive element could be done {\em after} reducing the inputs modulo $p,$
thus, without blowup, it is expensive; it introduces an $O(D^3)$ factor
into the overall complexity of the algorithm which is $O(D^2)$ otherwise.
We make some additional remarks about this in the conclusion.

\item Not invert ${\rm lc}(f_2)$, which can also cause a blowup, and can
also be more expensive than computing $g$.

\item Use rational reconstruction -- see \cite{Collins,MQRR,Pan,Wang}.
Otherwise a denominator bound would be necessary, but such bounds
are generally too large. The defect bound, usually the
(reduced~\cite{Bradford}) discriminant, which is part of the denominator
bound, is usually also too large.

\item Use trial division.  Otherwise we would need bounds for $H(g)$.
Such bounds will be a function of $f_1, f_2, L$ and will be much too
large when $g$ is small relative to $f_1$ and $f_2$, an important
special case.

\end{enumerate}
Encarnacion's paper confirms and deals with these items.
As a result, Encarnacion's algorithm is the fastest algorithm
for a single extension.
As for item 1, his paper deals only with a single extension, but he
does illustrate that modifying that extension (making $\alpha_1$
an algebraic integer) is not efficient. But if modifying one
extension $\alpha_1$ is not efficient, then modifying $n$ extensions
(replacing it by a primitive element) is certainly not efficient.

\subsection{Organization of the paper}
Our first goal is to generalize Encarnacion's algorithm to multiple
extensions without using a primitive element.
We do this in section 2 where we study the Euclidean algorithm in $L[x]$
modulo a prime $p$.

In section 2 we present our modular GCD algorithm and study its expected
time complexity.  We also describe how to modify the modular GCD algorithm so that it can
be used when one or more of the minimal polynomials defining
the number field $L$ are not irreducible and in section 4 we give
explicit code for how to do this in Magma.

In section 3 we present two implementations of our modular GCD algorithm,
one in Maple and one in Magma.  The data structure that we use for both
implementations, for representing polynomials and field elements, is
a recursive dense data structure.
We give details and explain why it is a good choice.

To demonstrate the effectiveness of the modular GCD algorithm in
$L[x]$ we compare it with several implementations of the Euclidean
algorithm over characteristic 0.  Based on the work of Maza and
Rioboo in \cite{Maza} we give a new primitive $\ZZ$-fraction-free
algorithm for $L[x]$ which is the best non-modular algorithm. Timing
comparisons comparing the two implementations of our modular GCD
algorithm with the various non-modular Euclidean algorithm based
implementations are given along with comparisons demonstrating the
effectiveness of the other improvements we have made.

\section{The modular GCD algorithm}
\input{primes}
\input{EAring}

\input{modgcd}

\subsection{When $L$ is not a field}
Until now we have assumed that $L$ is a field, i.e., we assumed
that $L_{i-1}$ is a field and each $m_i(z_i)$ is irreducible over $L_{i-1}.$
The algorithm does not verify these assumptions because testing
irreducibility of $m_i$ with a factorization algorithm could be costly,
and in many applications, it will be known a priori that
each $L_i$ is a field hence such tests would be redundant.
However, in the context of solving a systems of polynomial equations
over $\QQ$ with finitely many solution, Lazard in \cite{Lazard} presents
an algorithm for decomposing a lex Gr\"obner basis into a union of triangular sets
where univariate gcds are computed in $L[x]$ and $L$ is often not field,
that is, one or more of the $m_i$ are be reducible over $L_{i-1}$.
Another algorithm of Kalkbrenner in \cite{Kalkbrenner} also computes
gcds in $L[x]$ where $L$ is often not a field.
Kalkbrenner's algorithm decomposes a polynomial system into
a union of triangular sets using pseudo-remainders and gcd computations in $L[x].$
The problem of computing gcds efficiently in $L[x]$ when one or more of the
$m_i$ are reducible is studied by Maza and Rioboo in \cite{Maza}.
We will look at their algorithm in more detail in a later section.
As our algorithm is stated, if any $m_i(z_i)$ is reducible, and
the leading coefficient of a remainder in the Euclidean algorithm
(when run over $L$) is not invertible, our modular algorithm
will most likely enter an infinite loop because the
Euclidean algorithm mod $p$ will fail for all but finitely many $p$.
This is a serious flaw which we now address.

Let ${\bf d} = {\rm GCD}_L(f_1,f_2)$ be the output of
Euclidean algorithm over $L$ (over characteristic 0).
If ${\bf d} \neq$ ``failed'', then it is still true that all but
finitely many primes are good. In this case, the modular GCD algorithm
presented thus far will produce ${\bf d} \in L[x]$.
However, if ${\bf d}=$ ``failed'', then all but finitely many primes
are fail primes. So we can not expect the modular GCD algorithm to terminate.
We want to have a {\em modified} modular GCD algorithm that has
the following specifications:

\begin{enumerate}
\item It must always terminate, whether $L$ is a field or not.
\item If $L$ is a field, the output must be ${\rm GCD}_L(f_1,f_2) \in L[x]$.
\item If $L$ is not a field, then the output must one of the
following: Either the monic gcd in $L[x]$. Or the output
is ``failed'', in which case a second output must
be returned as well, namely a non-trivial factor $d_i$ of some $m_i,$
a zero divisor in $L_i$.
\end{enumerate}

\noindent
{\bf Example:} Let $L = \QQ(\alpha_1)$ where $m_1(z_1) = z_1^2-1$.
Let $f_1 = x^2 + \alpha$ and $f_2 = (\alpha+1) x + 1.$  Inverting
$\lc_x f_2 = \alpha+1$ will fail for all primes $p$.
Thus in our example the output of our modified algorithm
should be "failed", $z_1+1$.

\medskip
\noindent
{\bf Remark:}
Suppose $L$ is not a field and the Euclidean algorithm if run
in characteristic 0 would encounter a zero divisor.  The modification to
our modular algorithm described below will most probably output this zero
divisor.  It can, however, output a different zero divisor.

\medskip
\noindent
It is well known that the Euclidean algorithm can easily be modified
to meet the above specifications without calling a factoring algorithm:
The Euclidean algorithm ${\rm GCD}_L(f_1, f_2)$ in characteristic 0
will only fail if we divide by a zero divisor, that is, we try to
invert a zero divisor.  Inverses use the extended Euclidean algorithm
applied to $m_i(z_i)$ and some other element of $L_{i-1}[z_i]$ for some $i$.
The inverse only fails when this gcd is not 1, in which case a non-trivial
factor $d_i$ of $m_i$ has been found.  The {\em modified} Euclidean
algorithm will then return ``failed'' for the gcd of $f_1,f_2$,
but will also return $d_i(z_i)$ as second output.
Exactly how this is implemented will depend on the system.
In our Maple implementation, when we compute inverses in $L_i$ using
the extended Euclidean algorithm, if an inverse does not exist,
we generate a run-time error and return the non-trivial gcd
found as part of the error.
The calling routine may ``catch'' this error and process it.
In our Magma implementation, because Magma has no non-local goto
mechanism, we must use a different approach which we describe
in detail in the next section.

For efficiency reasons, we want to turn this into a modular algorithm.
If we run the modified Euclidean algorithm mod $p$,
using the same arguments as in lemma~\ref{finitep} one sees that for
all but finitely many $p$ the result will be ``failed'' with
$d_i$ mod $p$ as a second output.
So we make the following modification to the modular GCD algorithm:
In addition to all the steps done before, we will also store the
second outputs of the modified Euclidean algorithm mod $p$.
Each time the number of these second outputs reaches a certain threshold
(for example a Fibonacci number $F_n$) we combine them using Chinese remaindering,
apply rational reconstruction, and if rational reconstruction suceeds,
perform a trial division to see if we found a true factor $d_i \in L_{i-1}[z_i]$ of $m_i(z_i)$.
To prevent that a prime $p$, for which the second output is different
from $d_i$ mod $p$, can cause an infinite loop, we do not use
all available primes when computing $d_i$ with Chinese remaindering; instead
we omit the first $F_{n-2}$ primes, thus use only the last $F_{n-1}$ primes.

\section{Implementation}
At the end of this section we describe two implementations of our
modular GCD algorithm, one in Maple 9 \cite{Maple} and one in Magma 2.10 \cite{Magma}.
We give timing comparisons for the two implementations to demonstrate
the effectiveness of our improvements and for comparison with the Euclidean algorithm.

To fix notation, recall that $L = \QQ(\alpha_1,\ldots,\alpha_n$)
where $\alpha_i$ is algebraic over $L_{i-1} = \QQ(\alpha_1,\ldots,\alpha_{i-1})$,
and $m_i(z_i) \in L_{i-1}[z_i]$ is the minimal polynomial for $\alpha_i$ over $L_{i-1}$.
To implement the the modular GCD algorithm, we start with input polynomials
over $L$, reduce them modulo $p$ a machine prime so that they are over $L$ modulo $p$,
run the Euclidean algorithm retract them to be over $\ZZ$ for application of the
Chinese remainder theorem, reconstruct the rational coefficients so the output
is over $L$ and finally perform trial divisions over $L$.



\input{datastructure}
\input{trialdiv}
\input{maple}

\input{magma}

\input{maza}
\input{timings}

\input{conclusion}

\subsection*{Acknowledgment}
We acknowledge John Cannon and the Magma group for hosting
in Sydney in 2003 and Allan Steel for helping
us with our Magma implementation.
We also acknowlege Mark Moreno Maza for providing details of his implementation
of the fraction free algorithm in \cite{Maza}.

\end{document}

%% file: primes.tex
\subsection{lc-bad, fail, unlucky and good primes}
The {\em modular GCD algorithm} computes the {\em monic gcd}\ 
$g \in L[x]$ of $f_1$ and $f_2$.
It does this by reducing $f_1,f_2$ modulo one or more primes
and calling the {\em Euclidean algorithm mod $p$} for each of these primes $p$.
The modular GCD algorithm reconstructs $g$ from these modular images.
If the Euclidean algorithm mod $p$ outputs $g \bmod p$ we say $p$ is a good prime.
Only good primes should be used during the reconstruction for it to
be successful.
However, not all primes are good. We distinguish the following cases:
\begin{definition}
\label{definitionbad}
Let $f_1,f_2 \in L[x]$ and $g$ be their monic gcd.
We will distinguish four types of primes.
\begin{itemize}
\item {\em \bad primes}.
Let $m_1,\ldots,m_n$ be the minimal polynomials of the 
field extensions $\alpha_1,\ldots,\alpha_n$.
So $m_i(z)$ is a monic irreducible polynomial in 
$\Q(\alpha_1,\ldots,\alpha_{i-1})[z]$ and $m_i(\alpha_i)=0$.
If ${\rm den}(f_1)$, ${\rm den}(f_2)$ or
any leading coefficient of $\check{f}_2,\check{m}_1,\ldots,\check{m}_n$
vanishes mod $p$ then we call $p$ an {\em \bad prime}.

\item {\em Fail primes}.
If $p$ is not an \bad prime, and the Euclidean algorithm
mod $p$ returns ``failed'', then $p$ is called a {\em fail prime}.

\item {\em Unlucky primes}.
If $p$ is not an \bad prime nor a fail prime, and if the
output of the Euclidean algorithm mod $p$ has higher degree than $g$,
then $p$ is called an {\em unlucky prime}.

\item {\em Good primes}.
A prime $p$ is called a {\em good prime} if the Euclidean algorithm mod $p$
returns $g$ mod $p$.
Theorem~\ref{theorem1} in section~\ref{sectionEucl} says that all primes
that are not \bad are either fail, unlucky or good.
\end{itemize}
\end{definition}
{\bf Remarks:}
\begin{enumerate}
\item Our definition of \bad prime is not symmetric in $f_1,f_2$.
It could be that $p$ is \bad for $f_1,f_2$ but not \bad for $f_2,f_1$.
In that case, because of how we set up the algorithm,
we should either: not use $p$, or: interchange
$f_1,f_2$ mod $p$ before calling the Euclidean algorithm mod $p$.

\item Our definitions are not the same as the
standard definitions in~\cite{Brown}.
For example, it is possible that the Euclidean algorithm mod $p$ fails
even if the monic gcd of $f_1$ mod $p$, $f_2$ mod $p$ exists
and equals $g$ mod $p$.
We call such $p$ a fail prime and not a good prime.
This distinction is not necessary if $f_1,f_2 \in \Q[x]$
where there are no fail primes.

\item If $p \divides {\rm den}(g)$
(in the standard definition these primes are called bad primes)
then $g$ mod $p$ is not defined and so
$p$ can not be a good prime.
According to theorem~\ref{theorem1}, $p$ must then be either
\bbad, fail, or unlucky.

\item Minimal polynomials are monic so
the leading coefficients of $\check{m}_1, \ldots, \check{m}_n$ are
${\rm den}(m_1), \ldots, {\rm den}(m_n) \in \ZZ$.
However, ${\rm lc}(\check{f_2})$ is in general not an integer but
an algebraic number.

\item 
%
%
It is very easy to tell if a prime $p$ is \bad or not, but we can
not tell in advance if $p$ is fail, unlucky, or good.
So we will end up calling the Euclidean algorithm mod $p$ with fail, unlucky,
and good primes but never with \bad primes.
\end{enumerate}

\subsubsection{\bad primes}
If $f_1=5x+1$, $f_2=5x-1$ and $p=5$ then $p$ satisfies
our definition of an \bad prime as well as the definition of a good
prime. However, there are good reasons not to use any \bad prime.
Take for example $f_1=f_2=5x+1$.
Also, the proof of theorem~\ref{theorem1}
requires that $p$ not be \bbad.

Another example is $L=\Q(\alpha)$, $f_1,f_2 \in L[x]$ with
gcd $g = x + \alpha^3$,
$p=5$, and the minimal polynomial of $\alpha$ is
$m=z^5+z^4+\frac15 z^3 - \frac15$.
Because of preprocessing, in the algorithm we work with
$\check{m} = 5 z^5 + 5z^4+ z^3 - 1$. Modulo $p=5$ this becomes
$z^3 + 4$. If we used the prime $p=5$, it is easy to give an example $f_1,f_2$
where the Euclidean algorithm mod $p$ returns $g$ mod $(5,\alpha^3 + 4)$
which is $x+1$. But, viewing $\alpha$ as a variable,
$g \not\equiv x+1$ mod $5$.

For our algorithm, the best solution to the above problems is:
{\em never use  an \bad prime}.

\subsubsection{Fail primes}
Fail primes are primes for which the Euclidean algorithm mod $p$
tries to divide by a zero divisor, in which case it returns ``failed''.
Take for example $f_1 = x^2-1$, $f_2 = ax-a$ where $a = {2}^{1/5} + 5$.
Denote $a$ mod $p$ as $\overline{a}$.
The Euclidean algorithm mod $p$ will first try to make $f_2$ mod $p$
monic by multiplying it with $1/\overline{a}$.
But if $N(a)$, the norm of $a$, vanishes mod $p$
then $\overline{a}$ is zero or a zero-divisor, and the computation of
$1/\overline{a}$ fails. In this example $N(a) = 53 \cdot 59$
so the fail primes are $53$ and $59$.

The reason that in our terminology $53$ and $59$ are called fail primes
and not \bad primes in the example (after all, the problem was caused
by ${\rm lc}(f_2)$ mod $p$)
is to indicate how these primes are discarded:
We do not actively avoid these primes, instead, they
``discard themselves'' when the Euclidean algorithm mod $p$ is called.

One can also construct examples where $p$ is not \bbad,
${\rm lc}(f_2)$ is a unit mod $p$,
but $p$ still divides ${\rm den}(g)$  (occasionally such $p$
can be unlucky instead of fail).
Take for example $\alpha$ with minimal polynomial $m = z^3+3z^2-46z+1$,
$f_1 = x^3-2x^2+(-2\alpha^2 + 8\alpha +2 )x - \alpha^2 + 11\alpha - 1$,
$f_2 = x^3-2x^2-x+1$.
The monic gcd is $g = x - \frac{1}{91}\alpha^2- \frac{23}{91}\alpha - \frac{50}{91}$.
The denominator is ${\rm den}(g) = 91 = 7 \cdot 13$.
In this example, if $p \in \{7,13\}$ then $p$ is not \bad and the leading
coefficient of $f_2$ (as well as of $f_1$) is a unit mod $p$.
Nevertheless, $p$ can not be a good prime because $p \divides {\rm den}(g)$.
In this type of example $p$ must divide the discriminant.
For this reason, Encarnacion~\cite{Encarnacion} tests if the
discriminant is $0$ mod $p$ and avoids such primes.
However, even without the discriminant-test,
the primes $p \in \{7,13\}$ would still have been discarded at some point:
The Euclidean algorithm mod $p$
will calculate $r_3$ = $f_1$ mod $(p,f_2)$, try to make $r_3$ monic
and fail because the leading coefficient of $r_3$,
namely, $-2\alpha^2+8 \alpha+3$, is a zero divisor mod $p$.

Although one can generalize the discriminant-test
to $L$, 
our algorithm does not use it because it makes no difference for
the correctness of the algorithm.
For an intuitive explanation 
see lemma~\ref{lemmaTestNotNeeded} and for a proof see theorem~\ref{theorem1}.

\subsubsection{Unlucky primes}
Unlucky primes are not trivially detectable like \bad primes
and do not ``discard themselves'' like fail primes do, but
need to be detected and discarded nevertheless.
Fortunately, Brown~\cite{Brown} showed how to do this in
a way that is efficient and easy to
implement: Whenever modular gcd's do not have the
same degree, keep only those of smallest degree
and discard the others.

As an example, take $f_1 = x^2+(2\sqrt{5}+1)x+3$, $f_2 = x^2-x-1$,
$g = x + (\sqrt{5}-1)/2$. Then the Euclidean algorithm mod $2$
will return $x^2+x+1$, so $p=2$ is an unlucky prime.
But if $f_1 = x^2 + \sqrt{5}\ x + 1$, $f_2$ and $g$ the same as before,
then $p=2$ is a fail prime.

\subsubsection{Good primes}
\label{goodprimes}
All but finitely many primes must be good. This is because
if one would run the Euclidean algorithm in characteristic 0,
it would be a finite computation, and so there can only be finitely
many conditions on the primes and each condition only excludes
finitely many primes (see lemma~\ref{finitep}).


Of course we will not run the Euclidean algorithm
in characteristic 0, so this does not tell us which primes to use.
But this is not a problem because
to guarantee correctness of the
algorithm, just as in Brown's algorithm,
all we need to do is to avoid the \bad primes.
Experiments show that random primes are good with high probability. Hence,
even if there was an oracle that quickly provided good primes, it would not
noticeably improve the running time.

%% file: EAring.tex
\subsection{The Euclidean algorithm over a ring}
\label{sectionEucl}
Let $\alpha_1,\ldots,\alpha_n$ be algebraic numbers.
Let $L_i = \Q(\alpha_1,\ldots,\alpha_i)$ and $L=L_n$.
Let $\degr_i$ be the degree of $\alpha_i$ over
$L_{i-1}$.
The dimension of $L$ as a $\Q$-vector space is $\degr_* :=
\degr_1\cdots \degr_n$.
A basis of $L$ is:
\[
	M := \{ \prod_{i=1}^n \alpha_i^{e_i}\ \vert \ 0 \leq e_i < \degr_i \}.
\]
Let $\tilde{R}$ be the set of all $\Z$-linear combinations of $M$
and let $\tilde{R}_i = \tilde{R} \bigcap L_i$.
Let $\minp_i$ be the minimal polynomial of $\alpha_i$
over $L_{i-1}$. The degree of $\minp_i$ is $\degr_i$, $\minp_i$ is {\em monic} (the leading
coefficient is ${\rm lc}(\minp_i)=1$)
and $\minp_i(\alpha_i)=0$.
The coefficients of $\minp_i$ are in $L_{i-1}$.
Let $l_i$ be the smallest positive integer such that the coefficients
of $l_i \minp_i$ are in $\tilde{R}_{i-1}$.
Denote $\Fp = \Z/p\Z$ and $l_* = l_1\cdots l_n$.

In general $\tilde{R}$ is not a ring.
For example, $\alpha_1 \in \tilde{R}$, but $\alpha_1^{\degr_1}$ is not
in $\tilde{R}$ unless $l_1=1$.
When $a,b \in \tilde{R}$, to compute the product $ab \in L$ we
replace $\alpha_1,\ldots,\alpha_n$ by variables $z_1,\ldots,z_n$,
then multiply $a$,$b$
as polynomials, and after that
take the remainder modulo the polynomials $\minp_1(z_1),\ldots,\minp_n(z_n)$.
During this computation we only divide a bounded number of times by 
$l_1,\ldots,l_n$. Hence, if $k$ is a sufficiently large integer, then
$l_*^k ab \in \tilde{R}$ for all $a,b \in \tilde{R}$.

If $a \in L$ then define the {\em denominator} of $a$ as the smallest
positive integer ${\rm den}(a)$ such that ${\rm den}(a) a \in \tilde{R}$.
Note that $\tilde{R}$, and hence ${\rm den}(a)$, depends on the choice of
$\alpha_1,\ldots,\alpha_n$.
For example, if $\alpha_1 = \sqrt{8}$
and $a = \frac12 \alpha_1$ then ${\rm den}(a)=2$.
For $a \in L$ one has $a \in \tilde{R} \Longleftrightarrow {\rm den}(a)=1$,
in particular ${\rm den}(0)=1$.
Define
\begin{eqnarray}
	R_p & = &
	\{ a \in L\ \vert \ {\rm den}(a) \not\equiv 0 {\rm \ mod \ } p \} \\
	& = & \{ \frac{a}m\ \vert \ a \in \tilde{R},\ m \in \Z,\ m \not\equiv 0 {\rm \ mod \ }
p\}.
	\label{Rp}
\end{eqnarray}
If $a,b \in L$ then ${\rm den}(ab)$ divides ${\rm den}(a){\rm den}(b) l_*^k$
for some $k$. Hence, if $p \notdivides l_*$ then $R_p$ is a ring.
We will {\em always assume that $p$ does not divide $l_*$} so that $R_p$ is a ring
(if $p \divides l_*$ then $p$ is an \bad prime).
Denote
\[
	\Z_{(p)} = R_p \bigcap \Q = \{ \frac{a}{m}\ \vert\ a,m \in \Z, \ m
	\not\equiv 0 {\rm \ mod \ } p \}.
\]
Then $R_p$ is a $\Z_{(p)}$-module with basis $M$.
Define 
\[ \overline{R} = R_p/pR_p. \]
If $a \in R_p$ then we use the notation $\overline{a}$, or also $a$ mod $p$,
for the image of $a$ in $\overline{R}$. If $a \in L$, then
(primes that divide $l_*$ are always excluded)
\[
	\overline{a} {\rm \ is \ defined }
\Longleftrightarrow
	a \in R_p
\Longleftrightarrow
	p \notdivides \ l_* {\rm den}(a).
\]
If $\overline{a}$ is defined we will say that $a$ {\em can be reduced mod} $p$.

Now $\overline{R}$ is a ring and also an $\Fp$-vector space with
basis $M$ mod $p$. We can do the following identifications:
\begin{equation}
\label{tensorprods}
	    R_p = \tilde{R} \otimes_{\Z} \Z_{(p)}, \ \ \
	      L = \tilde{R} \otimes_{\Z} \Q, {\rm \ \ and \ \ }
  \overline{R}  = \tilde{R} \otimes_{\Z} \Fp
\end{equation}

If $a \in L$ then $a$ is a {\em unit} in $R_p$ if and only if both $a$ and $1/a$
are in $R_p$  (whenever we write $1/a$ it is implicitly assumed that $a \neq 0$).
This is equivalent to $p \notdivides l_*{\rm den}(a){\rm den}(1/a)$.
If $a \in L$ we will call $a$ a {\em unit} mod $p$ if $a \in R_p$
and $\overline{a}$ is a unit in $\overline{R}$. The following lemma shows
that these two notions are equivalent.

\begin{lemma}
\label{unitmodp}
Let $a \in R_p$. Then
$a$ is a unit in $R_p$
if and only if $\overline{a}$ is a unit in $\overline{R}$.
\end{lemma}
{\bf Proof:}
If $a$ is a unit in $R_p$ then $a$ and $1/a$ are in $R_p$, hence
$\overline{a}$ and $\overline{1/a}$ are defined,
and since $a \mapsto \overline{a}$ is a ring homomorphism $R_p \rightarrow \overline{R}$
one sees that
$\overline{1/a}$ is the inverse of $\overline{a}$. Hence
$\overline{a}$ is a unit in $\overline{R}$. \\
Conversely, assume $\overline{a}$ is a unit. Then $a \neq 0$
so we can take $b:=1/a \in L$. To finish the proof we
need to show that $b \in R_p$.
Take the smallest integer $k$ for which $c := bp^k \in R_p$.
Since $k$ is minimal, we have $\overline{c} \neq 0$
but then $\overline{a}\overline{c}$ is the product of a unit
and a nonzero element in $\overline{R}$ and hence nonzero.
But $\overline{a}\overline{c}$ equals $\overline{abp^k} = \overline{p^k}$
so $\overline{p^k} \neq 0$, hence $k=0$, so $b \in R_p$ and $a$
is invertible in $R_p$.
\\

If $f \in L[\x]$ then
{\em the denominator} ${\rm den}(f)$ is defined as
the smallest positive integer such that
${\rm den}(f)f \in \tilde{R}[\x]$.
Now $f \in R_p[\x]$ if and only if $p \notdivides {\rm den}(f) l_*$.
The polynomial $\overline{f}$ is the image of $f$ in $\overline{R}[\x]$, and
is defined if and only if $f \in R_p[\x]$, in which case we 
will say that $f$ {\em can be reduced} mod $p$.
Furthermore, if $f$ and $\overline{f}$ have the same degree
(when ${\rm lc}(f)$ is nonzero mod $p$)
then we will say that $f$ {\em reduces properly} mod $p$.
If $p$ is not an \bad prime it means
that $f_1,f_2$ can be reduced mod $p$, and that $f_2$ reduces
properly mod $p$.

Let $0 \leq i \leq j \leq n$ and $a \in L_j$.
Multiplication by $a$ is an $L_i$-linear map $\psi: L_j \rightarrow L_j$.
The {\em characteristic polynomial} ${\rm cp}^j_i(a) \in L_i[\x]$ of $a$ over
the extension $L_j:L_i$ is defined as the characteristic polynomial of
this linear map.
The {\em trace} ${\rm Tr}^j_i(a)$ of $a$ over $L_j:L_i$ is the trace of $\psi$
and the {\em norm} $N^j_i(a)$ of $a$ over $L_j:L_i$ is the determinant of $\psi$.
Whenever we do not mention the extension $L_j:L_i$ it is assumed to
be $L:\Q$ (so $i=0$ and $j=n$) in which case we
write ${\rm Tr}(a)$, $N(a)$, ${\rm cp}(a)$.
Now the {\em integral closure of $\Z$ in $L$} is
\[
	{\cal O} = \{ a \in L\ \vert\  {\rm cp}(a) \in \Z[x]\}.
\]
This is a ring (see \cite{Hecke}), and the elements of ${\cal O}$
are called the {\em algebraic integers} in $L$.
We will use the following notation for the integral closure of $\Z_{(p)}$
in $L$
\[
	{\cal O}_p = \{ a \in L\ \vert\ {\rm cp}(a) \in \Z_{(p)}[x]\}.
\]
Suppose $a \in L$ and $m = {\rm den}({\rm cp}(a))$.
Then by definition $a \in {\cal O}_p$ if and only if $m \not\equiv 0$ mod $p$.
The characteristic polynomial of $ma$ is in $\Z[x]$, hence $ma \in {\cal O}$
and hence
\begin{equation}
\label{Op}
	{\cal O}_p =
	\{\frac{a}m\ \vert\ a \in {\cal O}, \ m \in \Z, \ m \not\equiv 0
	{\rm \ mod \ } p\}.
\end{equation}

\begin{lemma}
\label{unit}
If $0 \leq i \leq j \leq n$ and $a \in {\cal O}_p \bigcap L_j$ then $a$ is a unit
in ${\cal O}_p$ if and only if $N^j_i(a)$ is a unit in ${\cal O}_p$.
In particular, $a \in {\cal O}_p$ is a unit if and only if
$N(a) \in \Q$ is a unit in $\Z_{(p)}$, in other words,
both numerator and denominator
of $N(a)$ are not divisible by $p$.
The same is also true for $R_p$.
\end{lemma}
{\bf Remark:} If $p \notdivides l_*$ then $R_p \subseteq {\cal O}_p$
and the lemma implies that if $a \in R_p$ and
$1/a \in {\cal O}_p$ then $1/a \in R_p$. \\[5pt]
{\bf Proof:} The $L_i$-linear map $\psi:L_j \rightarrow L_j$
that corresponds to multiplication by $a$ is defined over ${\cal O}_p$, i.e.
the entries of the matrix of $\psi$ are in ${\cal O}_p$. If $N^j_i(a)$,
the determinant of $\psi$,
is a unit in ${\cal O}_p$ then the matrix is invertible over ${\cal O}_p$.
So then $\psi^{-1}(1) \in {\cal O}_p$, so $1/a \in {\cal O}_p$.
Conversely, if $a$ is invertible in ${\cal O}_p$ then $\psi$ is an invertible
linear map, so its determinant must be a unit. \\
Now $N(a) = N^n_0(a) \in L_0 = \Q$ and $\Q \bigcap {\cal O}_p = \Z_{(p)}$ so the
second statement follows. The proof for $R_p$ is the same,
although as always $p$ must not divide $l_*$ so $R_p$ is a ring.
\\

Note that one can check if $a \in R_p$ is invertible, and if so, compute
its inverse, with linear algebra over $\Z_{(p)}$ or over its field of
fractions $\Q$. The matrix of the system to be solved is the matrix of $\psi$.
The same also holds for $\overline{a} \in \overline{R}$, whenever it is
invertible, its inverse can be computed with linear algebra over $\Fp$.
But instead of solving linear equations, we will use 
the extended Euclidean algorithm to calculate inverses in $\overline{R}$.
However, this can increase the number of fail primes because
the calculation can fail even if $\overline{a}$ is invertible.
This is not a serious problem because the number of fail
primes will still be finite (see section~\ref{goodprimes}).

In the following, let ${\cal R}$ be a commutative ring
with identity $1 \neq 0$.
For a univariate polynomial $f \in {\cal R}[\x]$
define ${\rm monic}(f)$ as follows:
If $f=0$ then ${\rm monic}(f)=0$. If $f \neq 0$ and if the
leading coefficient ${\rm lc}(f) \in {\cal R}$ of $f$ is a unit, then define
${\rm monic}(f) = {\rm lc}(f)^{-1}f$. If $f \neq 0$ and ${\rm lc}(f)$
is not a unit then define ${\rm monic}(f)$=``failed''.

If $f_1,f_2 \in {\cal R}[\x]$ then the
{\em monic gcd} is defined as a polynomial $g \in {\cal R}[\x]$ such
that $g={\rm monic}(g)$ and for every
polynomial $h$ one has:  $h\divides f_1$ and $h\divides f_2$ if and only if
$h\divides g$. It is easy to show that if a monic gcd of $f_1,f_2$ exists,
then it is unique.
The well-known {\em Euclidean algorithm} over
${\cal R}$ works as follows. \\

\noindent {\bf Euclidean algorithm}. \\
{\bf Input:} a list $(f_1,f_2)$ of two univariate polynomials with coefficients in ${\cal R}$. \\
{\bf Output:} Either a message ``failed'' or the monic gcd. \\[-15pt]
\begin{enumerate}
\item	Set $r_1 = f_1$, $r_2 = f_2$, $i = 2$.
\vspace*{-2mm}
\item	\label{monic1}
	If $r_2 = 0$ then set $r_1$ = monic$(r_1)$.
	If $r_1$ = ``failed'' then return ``failed''.\vspace*{-2mm}
\item	If $r_i = 0$ then return $r_{i-1}$.
\vspace*{-2mm}
\item	\label{monic2}
	Set $r_i = {\rm monic}(r_i)$.
	If $r_i$ = ``failed'' then return ``failed''. \vspace*{-2mm}
\item	Let $r_{i+1}$ be the remainder of $r_{i-1}$ divided by $r_i$.
\vspace*{-2mm}
\item	\label{laststep} Set $i = i+1$ and go back to Step 3.
\end{enumerate}
{\bf Remark on a shortcut:}
Suppose that $r_i$ in step~3 is a nonzero constant.
Some implementations of the Euclidean algorithm over a field will then take a
{\em shortcut}: stop the computation, the output is 1.
Over a ring {\em we should not use this shortcut}\ because that would invalidate
lemma~\ref{st_exist} below. This plays a role because our algorithm will
not test if $p$ divides the discriminant.
We may only use the shortcut if $r_i$ is a unit.
For $r_i \in \overline{R}$ we can test that efficiently
by computing $N(r_i)$ mod $p$ (see lemmas~\ref{unitmodp},\ref{unit}).
\\

Denote $\EuclAlg_{\cal R}(f_1,f_2)$ as the output of this algorithm.
If $\EuclAlg_{\cal R}(f_1,f_2) \neq$ ``failed'' then
the sequence of polynomials $r_1,\ldots,r_m$ with $r_{m-1} \neq 0$, $r_m = 0$,
is called the {\em monic polynomial remainder sequence} of $f_1,f_2$.
\begin{lemma}
\label{st_exist}
If $g = \EuclAlg_{\cal R}(f_1,f_2)$
and $g \neq$ ``failed'' then the ideal
$(r_{i-1}, r_{i}) = {\cal R}[x]r_{i-1} + {\cal R}[x]r_{i}$
remains the same during each step.
In particular $(f_1,f_2) = (g)$ which implies:
\begin{enumerate}
\item There exist $s,t \in {\cal R}[x]$ such
      that $g = sf_1 + tf_2$.
\item $f_1$ and $f_2$ are divisible by $g$.
\item $g$ is the monic gcd of $f_1$ and $f_2$.
\end{enumerate}
\end{lemma}
{\bf Proof}: When we make $r_i$ monic, we divide by a unit, which does not
change the ideal. In step~\ref{laststep} we increase $i$ so we must show
that $(r_{i-1},r_i) = (r_i,r_{i+1})$ which is clear because $r_{i+1}$ is
the remainder of $r_{i-1}$ modulo $r_i$.
Hence $(f_1,f_2)=(r_1,r_2) = (r_{m-1},r_m) = (g,0) = (g)$. So $g \in (f_1,f_2)$
which is part 1, $f_1,f_2 \in (g)$ which is part 2. Finally, every $h$ that
divides both $f_1$ and $f_2$ divides any element of $(f_1,f_2)$ in particular
it divides $g$. Since $g$ is monic it satisfies precisely the definition
of the monic gcd. \\

\noindent {\bf Remark:} If $\EuclAlg_{\cal R}(f_1,f_2) \neq$ ``failed'' then the
{\em extended Euclidean algorithm}, which calculates $s$ and $t$ as well as $g$
will not fail either. \\

Let ${\bf d} = \EuclAlg_{\cal R}(f_1,f_2)$ be the output of the Euclidean algorithm.
If all leading coefficients during the computation
are units then the algorithm succeeds, the monic gcd exists and
equals ${\bf d}=r_{m-1}$.
If there is no monic gcd in ${\cal R}[x]$
then ${\bf d}=$ ``failed''.
If a monic gcd $g$ does exist then it is not necessarily true that
the algorithm will find it; the output ${\bf d}$ is then either $g$ or ``failed''.
A situation where the output is ``failed'' even when a monic gcd exists
is given in the following lemma.
\begin{lemma}
\label{lemmaTestNotNeeded}
Suppose $p \notdivides l_*$ and $f_1,f_2 \in R_p[x]$. Then  $f_1,f_2 \in {\cal O}_p[x]$.
Suppose a monic gcd $g \in {\cal O}_p[x]$ exists and that $g \not\in R_p[x]$.
Then $\EuclAlg_{{\cal O}_p}(f_1,f_2)=$ ``failed''.
\end{lemma}
{\bf Proof:}
If $p \notdivides l_*$ then $\alpha_1,\ldots,\alpha_n \in {\cal O}_p$,
hence $R_p \subseteq {\cal O}_p$ so $f_1,f_2 \in {\cal O}_p[x]$.
Since $\EuclAlg_{R_p}(f_1,f_2) =$ ``failed'', when we run the
Euclidean algorithm over $R_p$ we will encounter a leading coefficient in $R_p$
that is not a unit in $R_p$. But according to the remark after lemma~\ref{unit},
if $a \in R_p$ is not a unit in $R_p$ then it is also not a unit in ${\cal O}_p$
and hence the algorithm fails over ${\cal O}_p$ as well.
\\

If the ring ${\cal R}$ in the Euclidean algorithm is a field $L$,
then the output is never ``failed'', so $\EuclAlg_L(f_1,f_2)$ is
always the monic gcd of $f_1,f_2 \in L[x]$.
\begin{lemma}
\label{finitep}
Suppose $f_1,f_2 \in L[\x]$ and $r_1,\ldots,r_m \in L[\x]$
is the monic polynomial remainder sequence. Let ${\rm lc}_1,\ldots,{\rm 
lc}_{m-1}$ in $L$
be the leading coefficients that we divided by in steps~\ref{monic1} 
and~\ref{monic2}.
For all but finitely many primes the following holds:
\begin{enumerate}
\item $f_1,f_2 \in R_p[\x]$, and ${\rm lc}_1,\ldots,{\rm lc}_{m-1}$ are units in $R_p$.
\item $r_1,\ldots,r_m \in R_p[\x]$ and $\overline{r_1},\ldots,\overline{r_m}$
is the monic polynomial remainder sequence of $\overline{f_1},\overline{f_2}$.
\item $p$ is a {\em good prime} which means:
The monic gcd of $\overline{f_1},\overline{f_2}$ exists, will be found
by the Euclidean algorithm, and equals $\overline{g}$ where $g \in L[x]$ is the monic
gcd of $f_1,f_2$.
\end{enumerate}
\end{lemma}
{\bf Proof:} Part 1 holds for all primes that
do not divide any of the following:
$l_*$, ${\rm den}(f_1)$, ${\rm den}(f_2)$, ${\rm den}({\rm lc}_i)$, ${\rm den}(1/{\rm lc}_i)$
for $i<m$. Since these are finitely many
integers, all nonzero, we see that part 1 holds for all but finitely
many primes.
The only divisions in the Euclidean algorithm are divisions
by ${\rm lc}_i$, so if the input is in $R_p[\x]$
and all ${\rm lc}_i$ are units in $R_p$, then all polynomials
in the $\EuclAlg_L(f_1,f_2)$ computation are in $R_p[\x]$.
Induction shows that
$\overline{r_1},\ldots,\overline{r_m}$
is precisely the monic polynomial remainder 
sequence of $\overline{f_1},\overline{f_2}$, so part 2 follows from part 1.
Part 3 follows from part 2.
\\

Since we will only run the Euclidean algorithm in
$\overline{R}[\x]$ for various primes $p$, and not in $L[\x]$,
we do not know the values of ${\rm lc}_i$.
So the lemma does not tell us which primes are good, it only
says that all but finitely many primes are good.
We now investigate the relation between
$\EuclAlg_{\overline{R}}(\overline{f_1},\overline{f_2})$
and $\EuclAlg_{L}(f_1,f_2)$ when $p$ is not an \bad prime.
\begin{theorem}
\label{theorem1}
Let $f_1,f_2 \in L[x]$ and let $g \in L[x]$ be the monic gcd.
Assume $p \notdivides l_* {\rm den}(f_1) {\rm den}(f_2)$, $f_2 \neq 0$
and ${\rm lc}(f_2) \not\equiv 0$ mod $p$, so $p$ is not an \bad prime.
Let ${\bf d} = \EuclAlg_{\overline{R}}(\overline{f_1},\overline{f_2})$.
If ${\bf d} \neq$ ``failed'' then
\[ {\rm deg}({\bf d}) \geq {\rm deg}(g). \]
Furthermore, if ${\rm deg}({\bf d}) = {\rm deg}(g)$ then $g$
reduces properly mod $p$ and ${\bf d} = \overline{g}$.
\end{theorem}
{\bf Remark:} 
The theorem says that if $p$ is not \bad
then $p$ is either fail, unlucky, or good.
This implies that if \bad primes are avoided then the modular GCD algorithm
is correct. \\[5pt]
{\bf Proof:}
${\rm lc}(f_2) \not\equiv 0$ mod $p$, so if we 
assume ${\bf d} \neq$ ``failed'' 
then ${\rm lc}(f_2)$ must be a unit mod $p$,
see step~4 in the Euclidean algorithm.
There exist (see lemma~\ref{st_exist}) $s_0, t_0 \in R_p[x]$ such that
\[  \overline{s_0} \overline{f_1} + \overline{t_0} \overline{f_2} = {\bf d}.  \]
Now take a monic polynomial ${\bf d}_0 \in R_p[x]$ such
that ${\bf d} = \overline{{\bf d}_0}$. Then we have
\[
	s_0 f_1 + t_0 f_2 \equiv {\bf d}_0 {\rm \ mod \ } p.
\]
We will apply {\em Hensel lifting} to increase the modulus $p$ to a
higher power of $p$.
Define (starting with $i=1$)
\[
  h_i =
	(s_{i-1} f_1 + t_{i-1} f_2 - {\bf d}_{i-1} )/p^i \in R_p[x]
\]
and let ${\bf \rm q}_i, {\bf \rm r}_i \in R_p[x]$ be the quotient
and remainder of $h_i$ divided by ${\bf d}_0$
(this division works because ${\bf d}_0$ is monic).
Then define
\[
	\tilde{s}_i = s_{i-1} - p^i {\bf \rm q}_i s_0, \ \ \ 
	\tilde{t}_i = t_{i-1} - p^i {\bf \rm q}_i t_0, \ \ \
	{\bf d}_i = {\bf d}_{i-1} + p^i {\bf \rm r}_i.
\]
Then
\[
	\tilde{s}_i f_1 + \tilde{t}_i f_2 \equiv {\bf d}_i {\rm \ mod \ } p^{i+1}.
\]
Now $\tilde{s}_i,\tilde{t}_i$ can have higher degrees than $s_{i-1},t_{i-1}$.
To remedy this, do the following.
For $j\in \{1,2\}$ denote $f_{j,d} \in R_p[x]$ as a polynomial whose modular
image equals $\overline{f_j}/{\bf d}$.
Take ${\bf \rm q}_i s_0$ mod $p$, and divide it by
$\overline{f_{2,d}} \in \overline{R}[x]$. This division works because the
leading coefficient of $\overline{f_{2,d}}$ is ${\rm lc}(f_2)$ mod $p$,
which is invertible.
Take $q,r \in R_p[x]$ such that $\overline{q},\overline{r}$ are the
quotient and remainder of this division. Take $q,r$ in such a way that
they have the same degree as $\overline{q},\overline{r}$.
Then define
\[
	s_i = s_{i-1} - p^i r, {\rm \ \ and \ \ }
	t_i = t_{i-1} - p^i({\bf \rm q}_i t_0 + q f_{1,d}),
\]
and we still have
\[
	s_i f_1 + t_i f_2 \equiv {\bf d}_i {\rm \ mod \ } p^{i+1}.
\]
We can now increase $i$ and do the next Hensel step, and continue in
this way.
Because ${\rm deg}(r) < {\rm deg}(\overline{f_{2,d}})$
and ${\rm deg}({\bf \rm r}_i) < {\rm deg}({\bf d}_0)$, the degrees
of $s_i$ and ${\bf d}_i$ will be bounded as $i$ increases, and hence the
degree of $t_i$ mod $p^{i+1}$ is bounded as well.
So when $i \rightarrow \infty$,  the limit $\hat{s},\hat{t},\hat{\bf d}$
of $s_i,t_i,{\bf d}_i$ exists in the ring $\hat{R}_p[x]$ defined
below. \\
Denote $\hat{\Z}_p$ as the ring of $p$-adic integers. $\hat{\Z}_p$ is
the completion of $\Z_{(p)}$ with respect to the $p$-adic valuation norm.
Let $\hat{\Q}_p$ be the field of $p$-adic numbers, the field of fractions
of $\hat{\Z}_p$.
Denote $\hat{L}_p = R_p \otimes_{\Z_{(p)}} \hat{\Q}_p = L \otimes_{\sQ} \hat{\Q}_p$.
This is in general not an integral domain because minimal polynomials can become
reducible when one replaces $\Q$ by a larger field $\hat{\Q}_p$.
Denote $\hat{R}_p =   
R_p \otimes_{\Z_{(p)}} \hat{\Z}_p$.
Now $\hat{R}_p$ and $L$ can be viewed as subrings of $\hat{L}_p$
and
\begin{equation}
	\label{intersection}
	R_p = \hat{R}_p \bigcap L
\end{equation}
After doing infinitely many Hensel steps we find
$\hat{s},\hat{t},\hat{\bf d} \in \hat{R}_p[x]$ such that
\[ \hat{s}f_1 + \hat{t}f_2 = \hat{\bf d}. \]
Now $\hat{\bf d}$ is monic and
$ {\rm deg}(\hat{\bf d}) = {\rm deg}({\bf d}_0) = {\rm deg}({\bf d}) $
because the $p^i {\bf \rm r}_i$, $i=1,2,\ldots$, that we added to ${\bf d}_0$
have smaller degree than ${\bf d}_0$.
The polynomials $f_1,f_2$ are elements of $L[x]g \subseteq \hat{L}_p[x]g$.
Hence $\hat{s}f_1 + \hat{t}f_2$, which equals $\hat{\bf d}$, is a also
an element of $\hat{L}_p[x]g$.
But $\hat{\bf d} \neq 0$ so
\[	  {\rm deg}({\bf d})
	= {\rm deg}(\hat{\bf d}) \geq {\rm deg}(g).
\]
If the degrees are the same then $\hat{\bf d} = g$
because $g$ is the only monic element of $\hat{L}_p[x]g$
of that degree.
Equation~(\ref{intersection}) then implies $g \in R_p[x]$
(recall that $\hat{\bf d} \in \hat{R}_p[x]$ and $g \in L[x]$).
So $g$ can be reduced mod $p$. Hence $g$ reduces properly
mod $p$ because it is monic.
The theorem now follows because ${\bf d}$ equals $\hat{\bf d}$ mod $p$,
which equals $g$ mod $p$.


%% file: modgcd.tex
\subsection{The Modular GCD Algorithm in $L$[x]}

We give a high-level description of the modular GCD algorithm.

\noindent

\bigskip
\noindent
{\bf Modular GCD algorithm.} \\
{\bf Input:} Non-zero $f_1,f_2 \in L[x]$, $L$ a number field. \\
{\bf Output:} $g$, the monic gcd of $f_1$ and $f_2$. \\[-16pt]
\begin{itemize}
\item[1.] Preprocessing: Set $n=0$, $f_1=\check{f}$ and $f_2=\check{f}_2$.
\vspace*{-2mm}
\item[2.] Main Loop: Take a new prime $p$ that is not lc-bad.
\vspace*{-2mm}
\item[3.] Let ${\bf d}$ be the output of the Euclidean algorithm applied
          to $f_1$ and $f_2$ mod $p$.
          If ${\bf d} =$``failed'' then go back to step 2.
\vspace*{-2mm}
\item[4.] If ${\bf d}=1$ then return $1$.
\vspace*{-2mm}
\item[5.] If $n=0$ or ${\rm deg}({\bf d}) < {\rm deg}(c)$ then \\
          set $c={\bf d}, m=p, n=1$ and go to step 8.
\vspace*{-2mm}
\item[6.] If ${\rm deg}({\bf d}) > {\rm deg}(c)$ then go back to step 2.
\vspace*{-2mm}
\item[7.] Let $c$ be the output of applying Chinese remaindering to $c$ mod $m$ and
          ${\bf d}$ mod $p$.  Set $m = mp, k = k+1$.
\vspace*{-2mm}
\item[8.] Apply rational reconstruction to obtain $h \in L[x]$ from $c$ mod $m$. \\
          If this fails, go back to step 2.
\vspace*{-2mm}
\item[9.] Trial division: If $h|f_1$ and $h|f_2$ then return $h$, \\
          otherwise, go back to step 2.
\end{itemize}

\medskip
\noindent
Step 1 is a preprocessing step.  We compute $\check{f}_1$ and $\check{f}_2$,
the semi-associates of $f_1$ and $f_2$ respectively, that is,
we cancel any rational scalar from the input polynomials before proceeding.
We do not compute $\tilde{f}_1$ or $\tilde{f}_2$, the monic
associates of $f_1$ and $f_2$ which can cause a blowup.

Since lc-bad and fail primes are actively discarded in steps 2 and 3,
the primes $p_1, p_2, ..., p_k$ remaining after step 6 are
either all unlucky or all good.  Let $m = \Pi_{i=1}^k p_i$.
Suppose rational reconstruction succeeds at step 8 with output $h.$
If $h|f_1$ and $h|f_2$ then $h = g$ and the modular GCD algorithm terminates.
If either trial division fails then from Theorem 1
either $m$ is not yet large enough to recover the rational
coefficients in $g$ or all primes are unlucky.
Before we state the time complexity of the algorithm
we examine three technical problems.

\subsubsection*{Problem 1: The Trial Divisions}
If $h \ne g$ the trial divisions $h|f_1$ and $h|f_2$ in step 9 may
be very expensive because the rational coefficients in the quotient
$f_1/h$ may be much larger in length than those in $f_1/g$.
There are many ways to engineer the algorithm so that this
happens with very low probability.

One is to modify the trial division algorithm so it first
tests if $h|f_1 \bmod q$ and $h|f_2 \bmod q$ for some prime $q$ before
attempting divisions in characteristic 0.  For this test to be of value
the prime $q$ must be different from the primes used thus far
by the modular GCD algorithm.  Magma, for example, reserves
a special prime not used by modular algorithms for this purpose.

A second way is to build into the rational reconstruction algorithm
some redundancy so that if it succeeds with output $h$ then $h = g$
with high probability. This is our preferred approach.  To do this
one can either modify Wang's rational reconstruction algorithm in
\cite{Wang81,Wang}, or use the algorithm of Monagan in \cite{MQRR}.

A third possibility is to modify the modular GCD algorithm
so that when rational reconstruction succeeds with output $h$,
we compute $g_{k+1}$, the GCD modulo an additional prime $p_{k+1}$
and require that $h \equiv g_{k+1} \bmod p_{k+1}$ before we
attempt the trial divisions.  Maple version 8, for example,
uses this approach for a number of modular algorithms.

\subsubsection*{Problem 2: Rational Reconstruction is not Incremental}

When we apply the Chinese remainder theorem to compute the new
value of $c$ in step 7 such that $c_{new} \equiv c_{old} \bmod m$ and
$c_{new} \equiv {\bf d} \bmod p$, we can do this incrementally, i.e.,
in $O( \log m )$ instead of $O( \log^2 m )$ time per integer coefficient,
using only classical algorithms for integer arithmetic as follows:

\medskip
\noindent
\hspace*{2mm}    Step 9: Chinese remaindering. \\
\hspace*{2mm}    Set $\Delta = \bf d - c \bmod p$.   \\  
\hspace*{2mm}    Set $i = m^{-1} \bmod p$. \\
\hspace*{2mm}    Set $v = i \Delta \bmod p$. \\
\hspace*{2mm}    Set $c = c + m v$. \\
\hspace*{2mm}    Set $m = m p, k = k+1.$ \\

\noindent
However, no incremental version of rational reconstruction is known.
If one uses the Euclidean algorithm (see section 3.2), rational reconstruction will
cost $O( \log^2 m )$ per coefficient.  Suppose $g = x + n/d$ and $|n|,|d|<M$.
If rational reconstruction were applied at each step it will
introduce an $O( \log^3 M )$ component per rational coefficient
into the modular GCD algorithm.
This can be reduced to $O( \log^2 M )$ without increasing
the asymptotic cost of the other components of the modular GCD
algorithm and without using fast arithmetic
if we perform rational reconstruction periodically.
For example, after $F = 1, 2, 3, 5, 8, 13, 21, 34, 55, ...$  primes.

In practice the cost of rational reconstruction is
usually much less than $O( \log^3 M )$ per coefficient and the
Fibonacci sequence is much too sparse on most data.
Suppose $g$ has $N$ rational coefficients that need to be
reconstructed. Suppose rational reconstruction is designed so that
it will fail with high probability when the input is the image of a
rational number which cannot be reconstructed because $m$ is not yet
large enough. Suppose also it remembers the monomial in $g$ where it
failed in the previous step so that it always starts with a
coefficient for which it previously failed. Then if rational
reconstruction is applied at every step, the expected total cost of
rational reconstruction, assuming classical integer arithmetic, is
$O( \log^3 M + N \log^2 M ),$ that is, $O( \log^3 M / N + \log^2 M
)$ per coefficient.

\subsubsection*{Problem 3: Computing Inverses in the Euclidean Algorithm}

In Step 3 the Euclidean algorithm is applied over $L$ modulo $p$ which
is not a field in general; it is a finite ring $L_p$ with zero divisors
in general.  The (monic) Euclidean algorithm, described in section 2,
needs to invert the leading coefficient of the divisor, an element of $L_p$.
Units in $L_p$ can be inverted using linear algebra in $O(D^3)$ arithmetic
operations in $\ZZ_p$ where $D = [L:\QQ]$ is the degree of $L$ over $\QQ.$
However this would introduce an $O(D^3)$ factor into the modular GCD algorithm.
Thus we prefer to apply the Euclidean algorithm to compute inverses in $L_p$
because it requires only $O(D^2)$ arithmetic operations in $\ZZ_p$.
However, if $L_p$ is not a field, the Euclidean algorithm may fail
to compute an inverse even when the inverse exists.
If this happens we will also call $p$ a fail prime.
Thus a prime $p$ is a fail prime if the Euclidean algorithm with
input $f_1$ and $f_2$ in $L[x]$ fails modulo $p$ where inverses
are computed in $L_p$ using the Euclidean algorithm.
Thus there are two sources of failure.  One is elements of $L$ which
are not invertible modulo $p$ and the other is units in $L_p$ which
are not invertible by the Euclidean algorithm.
It is not hard to see that the number of fail primes is finite.
Run the Euclidean algorithm in characteristic 0 to invert
elements of $L.$  The conditions on $p$ for which elements
of $L$ are not invertible when using the Euclidean algorithm
involve integers of finite length and hence the number
of fail primes for any given input $f_1$ and $f_2$ is finite.

\subsection{Time Complexity of the Modular GCD Algorithm}
We estimate the average asymptotic time complexity of our
modular GCD algorithm for $L[x]$.  We will not include the cost
of the trial divisions in our complexity estimate and we
will state the expected running time in terms of
$\check{m_1}, ..., \check{m_k}$, $\check{f_1}$ and $\check{f_2}$.

Let $D$ be the degree of the number field $L$ and let
$C = \log \, \max_{i=1}^k H( \check{m_i}(z) )$, that is, $C$
bounds the size of the largest coefficient appearing in the $\check{m_i}$.
Let $N = \max( \deg_x(f_1), \deg_x(f_2) )$, $n = \deg_x(g)$,
$M = \log \, \max( H(\check{f_1}), H(\check{f_2}) ),$
and let $m$ be the number of good primes needed to reconstruct $g$.
In most cases $m \in O(M)$ though it can happen that the coefficients
of $g$ are larger than those of $f_1$ and $f_2$.

We will assume that the probability that a prime is good is high
so that $m$ is close to the actual number of primes that were used.
This assumption is true in practice when we use 30 bit primes.
However, for theoretical completeness of the complexity estimate,
we would need to determine some ${\bf B} = B(f_1,f_2,L)$ such that
if $p > \log \bf B$ then the probability that $p$ is good is greater
than some constant, say $1/2$.
Moreover, we require that $B(f_1,f_2,L)$ is a polynomial function
of the size of $f_1,f_2,L$, i.e., polynomial in $D,C,N,M$.
We did not determine such $\bf B$ because it appears to be difficult
to obtain a useful result and secondly, this issue would
not have consequences for the algorithm in practice (one hardly ever
encounters primes that are not good).
However, we do claim that such a bound that is
polynomial in $D, C, N, M$ exists.

Because neither of our implementations use asymptotically
fast arithmetic throughout it makes sense for us to first assume
classical arithmetic, i.e., quadratic algorithms for integer
and polynomial arithmetic.  Under the assumptions made we have
\begin{theorem}
\label{theorem2}
The expected running time of our modular GCD algorithm is
\[
   O( m (C + M N) D + m N (N-n+1) D^2 + m^2 (n D + m) ) )
\]
arithmetic operations on integers of size $O( \log p )$ bits.
\end{theorem}
The three contributions are for reducing the minimal polynomials $m_1, ..., m_k$
and input polynomials $\check f_1$ and $\check f_2$ modulo $m$ primes (step 3),
applying the Euclidean algorithm $m$ times (step 3),
and reconstruction of $O(n D)$ rational coefficients (steps 7 and 8),
respectively.

The hardest gcd problems for our algorithm occur when $n = N/2 +
o(N)$ and when $m$ is large, that is, $m \in O(M).$  This happens
when the gcd $g$ and cofactors $\bar f_1$ and $\bar f_2$ are of
similar size. This is also when dividing $f_1$ and $f_2$ by $g$
using the classical division algorithm is most expensive. Under the
simplifying assumption that $C \le M N$, that is the coefficients of
the minimal polynomials are not larger than those in $f_1$ and
$f_2$, the expected time complexity for these ``hard'' gcds is $O(
M^2 (N D + M) + M N^2 D^2 ).$


%% file: datastructure.tex
\subsection{A Data Structure for $L[x]$ and $L_p[x]$}
Our Maple and Magma implementations both use a {\em recursive
dense} representation for polynomials. This is the representation
advocated by Stoutemyer in \cite{David} as the best overall
representation for polynomials based on the his system Derive.
We choose this data representation for elements of $L$ and for
polynomials in $L[x]$. That is we regard the inputs $f_1$ and $f_2$
as polynomials in $x$ and $z_1,\ldots,z_n$. 

In our Magma implementation, we are implicitly using this representation
as we construct $L[x]$ as a tower of univariate polynomial extensions over
$\QQ$. In Magma, univariate polynomials are represented as a vector
of coefficients, that is, a dense one-dimensional array of coefficients.
In our Maple implementation, we have implemented a recursive dense data type.
The datatype, implemented in Maple code, is being implemented in the Maple kernel.

We describe the Maple data type {\tt <poly>} using a BNF notation.
\begin{verbatim}
  <poly> ::= POLYNOMIAL( <ring>, <data> )
  <ring> ::= [ <char>, <vars>, <exts> ]
  <char> ::= <nonnegative integer>
  <vars> ::= vector(<variables>)
  <data> ::= <immediate integer> | <rational number> | vector(<data>)
  <exts> ::= vector(<data>)
\end{verbatim}

\noindent
The characteristic of the ring is encoded by \verb+<char>+
and \verb+<exts>+ is a vector of the minimal polynomials.
Thus the ring to which the polynomial belongs is encoded
explicitly in the data structure.  Since the ring information
is identical for polynomials in the same ring it is stored
once so that the cost of storing the ring information
is one word (a pointer) per polynomial.

The bottom of the data structure is a word of storage which is
either a pointer to a rational number or an immediate integer.
In Maple 9, on a 32 bit computer, immediate integers are signed
integers of 30 bits in length where one bit is used to distinguish
them from pointers.

%

In a recursive dense representation a zero coefficient at any level
in the data structure, except the bottom level, is represented by
the immediate integer 0 (or the nil pointer). This means that every
algorithm must treat 0 as a special case. This exceptional case does
not bother us because in the implementation of most operations, 0 is
a special case anyway. In the Maple examples below, vectors are
indicated by square brackets.

\medskip
\noindent
{\bf Example 1:} The representation of the polynomial
$z^4-10 z^2 + 1$ in characteristic 0 and characteristic 3 is
\begin{verbatim}
  POLYNOMIAL( [0,[z],[]], [1,0,-10,0,1] )
  POLYNOMIAL( [3,[z],[]], [1,0,2,0,1] )
\end{verbatim}
The empty vector \verb+[]+ indicates that there are no extensions and the
data in both these examples is a vector of machine integers.
Allowing one word as a header word for the POLYNOMIAL structure and for
each vector, the storage requirement for both polynomials is 16 words.
Since the ring information can be shared between polynomials over the same
ring, a more accurate count is that 9 words are required. From
now on we count 1 word (a pointer) for the ring storage.

\medskip
\noindent {\bf Example 2:} The representation for the polynomial
$x^2 - 3 z x + 5$ in $\QQ[x,z],$ $\QQ[z]/\langle z^2-2 \rangle [x],$
and $\ZZ_3[z]/ \langle z^2-2 \rangle [x]$ is
\begin{verbatim}
  POLYNOMIAL([0, [x, z], []], [[5], [0, -3], [1]])
  POLYNOMIAL([0, [x, z], [[-2, 0, 1]]], [[5], [0, -3], [1]])
  POLYNOMIAL([3, [x, z], [[1, 0, 1]]], [[2], 0, [1]])
\end{verbatim}
\noindent
The storage requirement is 14, 14 and 11 words respectively.

\medskip
\noindent {\bf Example 3:} Even for moderately sparse polynomials,
the recursive dense data structure is surprisingly compact. Consider
the sparse polynomial $1+2x^n+3y^n+4z^n.$ Our data structure for
this polynomial for $n=3$ is
\begin{verbatim}
  POLYNOMIAL(R, [[[1,0,0,4], 0, 0, [3]], 0, 0, [[2]]]);
\end{verbatim}
\noindent This is 24 words.  In general it is $15+3n$ words. One of
the main sparse representations for polynomials that is used in
AXIOM is a linked list of pairs where each pair is a pointer to a
coefficient and a pointer to a monomial where the monomial $x^i y^j
z^k$ would be stored as an exponent vector $[i,j,k]$.  Thus each
non-zero term of the polynomial requires $2+2+4=8$ words of storage.
On our example this would be 35 words, allowing 3 words for the top
level of the data structure. On this example, the recursive dense
representation uses less storage for $n \le 6$.

\medskip
\noindent
{\bf Example 4:} Multiple extensions are handled in the obvious way.
Consider the polynomial $x^2 - \sqrt{2}/3 \ x + \sqrt{3}/2.$
We show how to input this polynomial in two ways, first,
directly, using the {\tt rpoly} command which
converts from Maple's native sum-of-products
representation for formulae to the POLYNOMIAL data structure,
and secondly, by first creating the number field and
polynomial ring using the {\tt rring} command.
We then reduce the polynomial $g$ modulo $p=5$.
{
\begin{verbatim}
  > f := rpoly(x^2-u/3*x+v/2, [x,u,v], [u^2-2,v^2-3]);
                  2                          2       2
           f := (x  - 1/3 u x + 1/2 v) mod <u  - 2, v  - 3>
  > lprint(f);
    POLYNOMIAL([0, [x, u, v], [[[-2], 0, [1]], [-3, 0, 1]]],
    [[[0, 1/2]], [0, [-1/3]], [[1]]])
  > L := rring( [u,v], [u^2-2,v^2-3] );
            L := [0, [u, v], [[[-2], 0, [1]], [-3, 0, 1]]]

  > Lx := rring(L,x); # construct L[x] from L
          Lx := [0, [x, u, v], [[[-2], 0, [1]], [-3, 0, 1]]]

  > g := rpoly( x^2-u/3*x+v/2, Lx );
                  2                          2       2
           g := (x  - 1/3 u x + 1/2 v) mod <u  - 2, v  - 3>
  > h := phirpoly(g,5);
                                2            2       2
           h := (3 v + 3 u x + x ) mod <3 + u , 2 + v , 5>
\end{verbatim}
}

\medskip
\noindent An advantage of the recursive dense representation
is the following.  When we reduce mod $p$, using the {\tt
phirpoly} comand above, we obtain a recursive structure where the
bottom level of the structure, representing polynomials in
$\ZZ_p[v]$ in the example, is a vector of machine integers.
This is the most efficient representation for arithmetic in $\ZZ_p[v]$.
This is important because this is where most of the computation will
occur when the Euclidean algorithm is executed modulo $p.$

%% file: trialdiv.tex
\subsection{Trial Division}
Another bottleneck of the modular GCD algorithm is the trial divisions.
If $h$ is the result of rational reconstruction then we must check
that $h|f_1$ and $h|f_2$ to show that $h=g$.
Because these trial divisions can be expensive,
we have considered abandoning trial divisions altogether in favor of a
probabilistic result, that is, check that result of rational reconstruction
agrees, say, with the gcd modulo five additional primes instead of one.
However, in many applications where one computes gcd's, for example,
normalizing a rational function, one wants to compute
also the cofactors $f_1/g$ and $f_2/g$, hence,
the divisions cannot be avoided.

There are also situations where one cofactor is enough. If Trager's
factorization algorithm is used to factor a polynomial $f \in L[x]$
where $k = [L:\QQ]$, one computes $g_1 = \GCD(f,f_1)$ where $f_1$ is
an irreducible polynomial over $\QQ$ and $f_1$ is the norm of a
factor of $f$. Since the degree of $g_1$ is known to be $d = \deg
f_1 / k$ in advance, it is not hard to see that if the modular GCD
algorithm constructs a polynomial $h$ of degree $d$ and $h|f$ then
$h$ must also divide $f_1$ and hence $h = g_1$.
Since it is useful to compute the cofactor $f/g_1$ in Trager's
algorithm, but not the cofactor $f_1/g_1$, then the latter trial
division, which is usually the larger in degree, may be avoided.
This simple observation can make a significant improvement.

When dividing $f_1$ and $f_2$ by $h$ over $L$ using the classical
division algorithm, a very significant improvement can be obtained
if one avoids fractions as much as possible.  This idea of avoiding
fractions has been used to speed up many computations in computer
algebra. Notice that the leading coefficient of $\check h$ in the
modular GCD algorithm is an integer.  If also $l_i = {\rm den}(m_i)
= 1$, which is often the case, then the entire division algorithm
can be completed using only integer arithmetic. If $l_i \ne 1$ for
some $i$ then the division algorithm can still be modified to avoid
fractions. We show how to do this for univariate polynomials with
one field extension with minimal polynomial $M$.
\\

\noindent
{\bf Algorithm Fraction Free Long Division.} \\
{\bf Input:} $A,B \in \QQ[x,z], ~M \in \ZZ[z]$ : $B \ne 0,$ $\lc_x B \in \QQ$, and
       $\deg M \ge 1$. \\
{\bf Output:} $Q = A/B \bmod M$ if $B|A \bmod M$; ``failed'' otherwise.
\begin{itemize}
\item[]    Set $m = \deg_x A,$ $n = \deg_x B$ and $d = \deg_z M$.
\vspace*{-2mm}
\item[]    Set $i_a = {\rm ic}(A)$ and $a = A/i_a.$
\vspace*{-2mm}
\item[]    Set $i_b = {\rm ic}(B)$ and $b = B/i_b.$
\vspace*{-2mm}
\item[]    Set $l_b = \lc_x b$ and $l_m = \lc_z M$. Remark: $l_b, l_m \in \ZZ.$
\vspace*{-2mm}
\item[]    Set $s = 1$, $r = a$, and $q = 0$.
\vspace*{-2mm}
\item[]    While $r \ne 0$ and $m \ge n$ do
\vspace*{-2mm}
\begin{itemize}
\item[]        Set $l_r = \lc_x r$.  Remark: $l_r \in \ZZ[z]$.
\vspace*{-1mm}
\item[]        Set $g = \GCD( {\rm ic}(l_r), l_b ).$
\vspace*{-1mm}
\item[]        Set $s = (l_b/g) \times s$.
\vspace*{-1mm}
\item[]        Set $t = (l_r/g) \times x^{m-n}.$
\vspace*{-1mm}
\item[]        Set $q = q + t/s$.
\vspace*{-1mm}
\item[]        Set $r = (l_b/g) \times r - t \times b$.
\vspace*{-1mm}
\item[]        Set $k = \deg_z r$.
\vspace*{-1mm}
\item[]        While $r \ne 0$ and $k \ge d$ do
\vspace*{-0mm}
\begin{itemize}
\item[]            Set $l_r = \lc_z r$.  Remark: $l_r \in \ZZ[x]$.
\vspace*{-0mm}
\item[]            Set $g = \GCD( {\rm ic}(l_r), l_m ).$
\vspace*{-0mm}
\item[]            Set $t = (l_r/g) z^{k - d}.$
\vspace*{-0mm}
\item[*]            Set $r = (l_m/g) \times r - t \times M$.
\vspace*{-0mm}
\item[]            Set $s = (l_m/g) \times s$ and $k = \deg_z r.$
\vspace*{-0mm}
\end{itemize}
\item[]        Set $m = \deg_x r$.
\vspace*{-1mm}
\end{itemize}
\item[]    If $r \ne 0$ then output ``failed''.
\vspace*{-2mm}
\item[]
Set $Q = (i_a/i_b) \times q$ and output $Q$.
\end{itemize}

\noindent
The algorithm first makes the inputs $A$ and $B$ primitive over $\ZZ$.
We claim that each time round the outer loop and also each time
round the inner loop the following invariant holds:
$a \equiv b q + r/s (\bmod M)$ where $s \in \ZZ$ and $r$ has
integer coefficients.  From this the correctness of the
algorithm follows easily.
The outer loop reduces the degree of the remainder $r$ in $x$.
In the outer loop we multiply $r$ by the smallest possible integer
so that $\lc_x r$, a polynomial in $\ZZ[z]$, will be exactly
divisible by the integer $\lc_x b.$
The inner loop then reduces the remainder $r$ modulo $M$.
In the inner loop we multiply $r$ by the smallest integer
so that $\lc_z r$, a polynomial in $\ZZ[x]$, will be divisible by
the integer $\lc_z M$.
The scalar $s$, an integer, keeps track of the integer factors
of $\lc_x b$ and $\lc_z M$, respectively, that $r$ was multiplied
by so that the terms of quotient $q$ may be correctly computed.

\medskip
\noindent
{\bf Remark:} The algorithm works over any integral domain $D$ for which
GCDs exist.  That is, replacing $\ZZ$ by $D$ and $\QQ$ by the quotient
field $D/D$ generalizes the algorithm to work for inputs
$A,B \in (D/D)[z][x]$ and $M \in D[z]$.  One application of this is
for $D = \ZZ_p[t]$ which arises when computing a gcd over an algebraic
function field in a single parameter $t$ using a modular GCD algorithm.
There for each prime $p$ used, the algorithm makes trial divisions
in $\ZZ_p(t)[z][x]$ modulo $M(t)$.


%% file: maple.tex
%
%
%

\subsection{Maple Implementation}
Program {\tt NGCD}, our Maple implementation of our modular GCD
algorithm uses the recursive dense polynomial data structure
described in section 4.1. Here we demonstrate it's usage on three
problems. The first gcd problem is in $K[x]$ where $K=\QQ(\sqrt 2,
\sqrt 3).$  We create the field as $K = \QQ[a,b]/\langle a^2-2,b^2-3
\rangle,$ create the polynomial ring $K[x]$, convert the two given
polynomials $f_1$ and $f_2$ below from Maple's native sum of product
representation for polynomials to the recursive dense representation
described in section 4.1, and then compute and display their gcd
using the command {\tt NGCD.}  This command also prints some
diagnostic information.
{ \small
\begin{verbatim}
> read recden; read NGCD; read PGCD;
> K := rring([a,b],[a^2-2,b^2-3]);
          K := [0, [a, b], [[[-2], 0, [1]], [-3, 0, 1]]]
> Kx := rring(K,x):
> f1 := rpoly(x^2+(a*b-a-1)*x-a*b-2*b,Kx):
> f2 := rpoly(x^2+(a*b-4*a+1)*x+a*b-8*b,Kx):
> NGCD(f1,f2);
  NGCD: GCD in Q[a, b][x] mod <b^2-3, a^2-2>
   NGCD: Prime 1 = 46273
   NGCD: Prime 2 = 46271
  NGCD: Trial divisions over Z starting after 2 primes
                                  2       2
                  (a b + x) mod <b  - 3, a  - 2>
\end{verbatim}
}

\noindent We now demonstrate our implementation on two gcd problems
over $L = K[c]/\langle c^2-6 \rangle$ which is not a field. In the
first problem an error is generated. The error message shows the
zero divisor found in characteristic 0, namely, $c-ab$ and the
corresponding extension polynomial $c^2-6$ that it divides.

{ \small
\begin{verbatim}
> L := rring(K,c,c^2-6):
> Lx := rring(L,x);
     [0, [x, c, b, a], [[[[-6]], 0, [[1]]], [[-3], 0, [1]], [-2, 0, 1]]]
> f1 := rpoly(x^2+a*b*x+1,Lx):
> f2 := rpoly((c-a*b)*x+1,Lx):
> NGCD(f1,f2);
  NGCD: GCD in Q[c, b, a][x] mod <a^2-2, b^2-3, c^2-6>
   NGCD: Prime 1 = 46273
   NGCD: fail prime 46273
  Error, (in NGCD) zero divisor found, c^2-6, c-a*b
\end{verbatim}
}

\noindent The second example shows a gcd computation of two
polynomials in $L[x,y,z]$ succeeding even though $L$ is not a field.
Note that our {\tt NGCD} returns the primitive associate of the
monic gcd, that is, $\tilde{g} = {\rm den}(g) g.$
{ \small
\begin{verbatim}
> P := rring(L,[x,y,z]): # create L[x,y,z]
> f1 := rpoly((2*x+c*y+a*b+2*z)*(x-a*y*z-c)^2,P):
> f2 := rpoly((2*x+c*y+a*b+2*z)*(y-c*x*z-b)^2,P):
> NGCD(f1,f2);
  NGCD: GCD in Q[c, b, a][x, y, z] mod <-2+a^2, -3+b^2, c^2-6>
   NGCD: Prime 1 = 46273
   NGCD: Prime 2 = 46271
  NGCD: Trial divisions over Q starting after 2 primes
                                         2            2        2
           (a b + 2 z + c y + 2 x) mod <c  - 6, -3 + b , -2 + a >
\end{verbatim}
}

%% file: magma.tex
\subsection{Magma Implementation}


Here we give details and examples of a Magma implementation of our
modular GCD algorithm for polynomials over a number fields. The
algorithm cannot, in fact, be implemented in Magma 2.9. We will
describe modifications made by Allan Steel to Magma 2.10 that permit
our algorithm to be implemented.

In Magma, before one may compute with $f \in K[x]$, $K$ a number
field, the user must explicitly construct the number field $K$ and
the polynomial ring $K[x]$. In the following Magma
session\footnote{Lines beginning with the $>$ character are input
lines and other lines are Magma output} we construct $\QQ[z],$ input
the polynomial $m = z^2-2 \in \QQ[z],$ compute $m^2,$ construct $K =
\QQ(a) = \QQ[z]/\langle z^2-2 \rangle$ using the {\tt NumberField}
constructor, and then compute $a^3$.

{ \small
\begin{verbatim}
  > Q := RationalField();
  > P<z> := PolynomialRing(Q); // construct Q[z]
  > m := z^2-2;
  > m^2; // compute and display m^2
  z^4 - 4*z^2 + 4
  > K<a> := NumberField(m);
  > a^3;
  2*a
\end{verbatim}
}

\noindent Number fields may also be constructed with the quotient
ring constructor {\tt quo}. Our modular GCD algorithm supports both.
Magma users are more likely to use {\tt NumberField} because the
Magma library for it is extensive. The {\tt NumberField} constructor
requires, naturally, that minimial polynomials are irreducible
whereas the quotient ring constructor does not. As an example we
construct the number field $K = \QQ(\sqrt 2, \sqrt 3)$ and the ring
$L = K(\sqrt 6)$ using both approaches.

{ \small
\begin{verbatim}
  > P<v> := PolynomialRing(K);  K<b> := NumberField(v^2-3);
  > a^3/b^3; // computes sqrt(2)^3/sqrt(3)^3
  2/9*a*b
  > R<w> := PolynomialRing(K);
  > L<c> := NumberField(w^2-6);
                       ^
  Runtime error in 'NumberField': Argument 1 is not irreducible
\end{verbatim}
\begin{verbatim}
  // using the quotient ring constructor ...
  > P<u> := PolynomialRing(Q);  K<a> := quo<P|u^2-2>;
  > P<v> := PolynomialRing(K);  K<b> := quo<P|v^2-3>;
  > R<w> := PolynomialRing(K);  L<c> := quo<R|w^2-6>;
  > c^3*a^3/b^3;
  4/3*a*b*c
\end{verbatim}
}

\noindent
We create two polynomials $f_1$ and $f_2$ in $K[x]$ and
compute their gcd.  First we use the built-in {\tt Gcd} command
which uses the ordinary Euclidean algorithm and then we use {\tt modgcdA},
our modular GCD algorithm which prints the primes used (30 bit primes).

{ \small
\begin{verbatim}
  > P<x> := PolynomialRing(K);
  > f1 := x^2+(a*b-a-1)*x-a*b-2*b;
  > f2 := x^2+(a*b-4*a+1)*x+a*b-8*b;
  > Gcd(f1,f2);
  x + a*b
  > modgcdA(f1,f2);
  prime=1073741789
  prime=1073741783
  x + a*b
\end{verbatim}
}

\noindent
To implemement our modular GCD algorithm we need to compute over $K$ modulo
a prime $p$.  In our example this means we need to compute over the
finite ring $(\ZZ_p[u]/ \langle u^2-2 \rangle )[v]/ \langle v^2-3 \rangle$.
We may construct this ring in Magma as a composition of univariate
quotients using the {\tt quo} constructor.  Below we do this for $p=7$ and
then attempt to compute the ${\rm gcd}(f_1,f_2) \bmod p$
using Magma's {\tt Gcd} command.

{ \small
\begin{verbatim}
  > Z7 := GaloisField(7);
  > R7<z> := PolynomialRing(Z7);  K7<a> := quo<R7|z^2-2>;
  > R7<y> := PolynomialRing(K7);  K7<b> := quo<R7|y^2-3>;
  > P7<x> := PolynomialRing(K7);
  > f1 := x^2+(a*b-a-1)*x-a*b-2*b;
  > f2 := x^2+(a*b-4*a+1)*x+a*b-8*b;
  > Gcd(f1,f2);
       ^
  Runtime error in 'Gcd': Algorithm is not available for this kind
  of coefficient ring
\end{verbatim}
}

\noindent
The error arises because Magma refuses to execute the Euclidean
algorithm here because $K7$ is not a field.  So we attempt to
implement the (monic) Euclidean algorithm (from section 2) directly.

{ \small
\begin{verbatim}
 > r1 := f1 mod f2; r1; // f2 is already monic
 (3*a + 5)*x + (5*a + 6)*b
 > u := LeadingCoefficient(r1);
 > r1 := u^(-1)*r1; // make r1 monic
          ^
 Runtime error in '^': Argument is not invertible
\end{verbatim}
}

\noindent When a zero divisor is encountered, an error occurs, which
is expected because $p=7$ is, in fact, a fail prime. For the modular
GCD algorithm we would like to ``catch'' this error, compute the
zero divisor over $\ZZ_p$, and move on to the next prime, which is
what we do in our Maple implementation.  Unfortunately there is no
non-local goto facility in Magma.  A consequence of this is that our
modular GCD algorithm cannot be implemented in Magma 2.9 without
programming our own polynomial arithmetic operations from scratch.
In Magma 2.10, Allan Steel has implemented {\tt
IsInvertible}\footnote{Because the implementation of IsInvertible
uses the extended Euclidean algorithm, it may output false even
though the input is invertible in the ring.} for rings in Magma, in
particular for quotient rings, so that we can detect a zero divisor
before dividing by it. For example:

{ \small
\begin{verbatim}
  > IsInvertible(3*a+5);
  false
  > IsInvertible(3*a+4);
  true 5*a + 5
\end{verbatim}
}

\noindent This enables the following implementation of the Euclidean
algorithm for $f_1, f_2 \in R[x]$ where $R$ is a univariate quotient
ring over a field, to detect a zero divisor, and if a zero divisor
occurs, to compute it by calling the same algorithm recursively. Our
implementation outputs a pair of values. The output $(true,g)$ means
the algorithm succeeded and $g$ is the GCD($f_1,f_2$) in $R[x]$. The
output $(false,g)$ means the algorithm failed and $g$ is the zero
divisor in $R$ that the Euclidean algorithm encountered.

{ \small
\begin{verbatim}
  > forward GetZeroDivisor;
  > EuclideanAlgorithm := function(f1,f2)
  >    // Input f1,f2 in R[x], R a univariate quotient ring
  >    while Degree(f2) ge 0 do
  >       u := LeadingCoefficient(f2);
  >       t,i := IsInvertible(u);
  >       if not t then return false, GetZeroDivisor(u); end if;
  >       f2 := i*f2; // make f2 monic
  >       r := f1 mod f2; f1 := f2; f2 := r;
  >    end while;
  >    u := LeadingCoefficient(f1);
  >    t,i := IsInvertible(u);
  >    if not t then return false, GetZeroDivisor(u); end if;
  >    return true, i*f1;
  > end function;
\end{verbatim}
\begin{verbatim}
  > GetZeroDivisor := function(u)
  >    K := Parent(u);  // K = R[z]/<m>, m in R[z]
  >    m := Modulus(K); // m is in R[z]
  >    P := Parent(m);  // P = R[z]
  >    f := P!u; // this coerces u in K to R[z]
  >    t,g := EuclideanAlgorithm(m,f);
  >    if not t then return g; end if;
  >    return K!g; // coerces g in R[z] back to K
  > end function;
\end{verbatim}
}

\noindent
The example below shows the Euclidean algorithm hitting
the zero divisor $a+4$ in the subring $K7(a) = \ZZ_7[u]/\langle u^2-2 \rangle$
where note $u^2-2 = (u+3) (u+4)$.

{ \small
\begin{verbatim}
  > EuclideanAlgorithm(f1,f2);
  false a + 4
\end{verbatim}
}

\noindent
We now demonstrate our implementation on the two gcd problems
from the previous section, namely, over $L = K(\sqrt 6)$ which
is not a field.  In the Magma session below we construct $L = K(c)$
as $K[w]/\langle w^2-6 \rangle$ where $a=\sqrt 2$, $b = \sqrt 3$, and $c = \sqrt 6$.
The first gcd problem in $L[x]$ is for $f_1 = x^2+1$, $f_2 = (c-ab) x + 1$
where the $c - a b$ is not invertible.
Our algorithm correctly computes and outputs $w-a b$ a divisor
of $w^2-6$, the minimal polynomial for $L$.

{ \small
\begin{verbatim}
  > P<x> := PolynomialRing(L);
  > f1 := x^2+a*b*x+1;
  > f2 := (c-a*b)*x+1;
  > modgcdA(f1,f2);
  prime=1073741789
  prime failed
  hit zero divisor w - a*b
\end{verbatim}
}

\noindent
The second gcd problem is a multivariate gcd problem in $L[x,y,z]$.
Note that we convert the flat multivariate polynomial representation used for input
to a recursive univariate tower $L[x][y][z]$ to improve
the efficiency of the modular GCD algorithm.

{ \small
\begin{verbatim}
  > R<x,y,z> := PolynomialRing(L,3);
  > f1 := (2*x+c*y+a*b+2*z)*(x-a*y*z-c)^2;
  > f2 := (2*x+c*y+a*b+2*z)*(y-c*x*z-b)^2;
  > modgcdA(f1,f2);
  prime=1073741789
  prime=1073741783
  x + 1/2*c*y + z + 1/2*a*b
\end{verbatim}
}


%% file: maza.tex
%
%
%

\subsection{Triangular Sets}

In the following subsection we will make a timing comparison
comparing our modular GCD algorithm with the monic Euclidean
algorithm on polynomials in $L[x]$. The bottleneck of the monic
Euclidean algorithm is the many integer gcds that are computed to
add, subtract and multiply the fractions that appear. Arithmetic
with fractions can be reduced by using $\ZZ$-fraction-free
algorithms for arithmetic in $L$ and could be eliminated entirely if
one uses a $\ZZ$-fraction-free GCD algorithm for $L[x]$. In order to
properly demonstrate the superiority of the modular GCD algorithm we
want to include in our timing comparison an implementation of the
best possible non-modular GCD algorithm for $L[x].$ The
fraction-free algorithm of Maza and Rioboo \cite{Maza} for computing
a gcd over a triangular set applies to our problem. We modify the
ideas of Maza and Rioboo to construct a {\it primitive}
$\ZZ$-fraction-free GCD algorithm for $L[x]$ which, based on our
experiments, is clearly the fastest of the non-modular algorithms
for gcds in $L[x]$.



\subsubsection*{Triangular sets}

For $1 \le i \le n$ let $P_i=\QQ[z_1,z_2,...,z_i],$ $m_i \in P_i,$ and
$T_i = \langle m_1, ..., m_i \rangle,$ $K_i = P_i/T_i$ and
let $T = T_n$, $P = P_n$ and $K = K_n.$
It is clear that $K_i$ is isomorphic to $L_i$ and thus a
gcd computation in $L_i[x]$ is equivalent to a gcd computation in $K_i[x]$.
The set of generators $m_1, ..., m_n$ for $T$ is called a {\it triangular set}
because $m_i$ is a polynomial in $z_1, ..., z_i$ only for $1 \le i \le n$.
Such sets arise naturally in elimination algorithms and in that context it will
often be the case that one or more of the $m_i$ are reducible over $K_{i-1}$
and thus $K$ is not a field in general.


In \cite{Maza}, Maza and Rioboo show how to compute $g$ modulo $T$ by modifying the
subresultant gcd algorithm for $K[x]$ to be fraction-free, that is, to work
inside the ring $\ZZ[z_1,z_2,....,z_n][x].$ Their algorithm outputs either
an associate of the gcd of $f_1,f_2$ or it outputs a non-trivial factor
of some $m_i.$  Their algorithm works if $\ZZ$ is replaced by any integral
domain where gcds exist.  In the context of polynomial systems
this would apply if there were parameters in the system.
For simplicity of exposition, let us suppose that for $1 \le i \le n$,
$m_i \in P_i[z_i]$ is monic over $\ZZ$, that is, ${\rm den}(m_i) = 1,$
so that reduction modulo $T$ does not introduce fractions.
And let us assume for the moment that $K$ is a field.
We recall the notion of pseudo division in $K[x].$

\begin{definition}
Let $f, g \in K[x]$ be non-zero, $\delta = \deg f - \deg g + 1 > 0,$
$c = {\rm lc}(g)$ and $\mu = c^{\delta}$.
The {\it pseudo-remainder} and {\it pseudo-quotient} of $f$ divided by $g$
are the polynomials $\bar r$ and $\bar q$, respectively, satisfying
$\mu a = b \bar q + \bar r$ and $\bar r = 0$ or $\deg \bar r < \deg b$.
\end{definition}
The key observation about pseudo-division is that if $f$ and $g$ have
no fractions on input, that is, ${\rm den}(f) = {\rm den}(g) = 1,$ and the
usual division algorithm is applied to $\mu f$ divided by $g$, no fractions
appear in the division algorithm and ${\rm den}(\bar r) = {\rm den}(\bar q) = 1$.

Maza and Rioboo define the notion of a quasi-inverse for commutative
rings with identity.  We specialize the definition to $K.$
\begin{definition}
Let $u \in K.$  Then $v \in K$ is a
{\it quasi-inverse} of $u$ if ${\rm den}(v) = 1$ and $u v = r$
for some integer $r.$
\end{definition}
\noindent
{\bf Example:} Let $u \in K = \QQ[z]/\langle m \rangle$ where $m$ is monic and
irreducible with ${\rm den}(m) = 1$.  If ${\rm den}(u) = 1$ then there exist
$s,t \in \ZZ[z]$ such that $s m + t u = r$ where $r \in \ZZ$ is the resultant
of $m$ and $x$.  Thus $v = t$ is a quasi-inverse of $u$ in $K$.
The polynomials $s,t$ and resultant $r$ can be computed without any fractions
using the extended subresultant algorithm.

\bigskip
\noindent
{\bf Remarks:} The definition for quasi-inverse is unique up to multiplication
by a non-zero integer and an algorithm for computing a quasi-inverse of $u$ may
or may not return the quasi-inverse of $u$ with smallest positive $r$.
Notice that in the case where $d = {\rm den}(u) > 1,$ if $v$ is a quasi-inverse
for $d u,$ then $d v$ is a quasi-inverse for $u.$

\bigskip
Let us assume for now that we know how to compute a quasi-inverse of $u \in K$.
In the monic Euclidean algorithm for $K[x]$ (see section 2) we make $r_i$ monic,
that is, we multiply $r_i$ by $u^{-1}$ where $u = {\rm lc}(r_i)$.  To obtain
a $\ZZ$-fraction-free algorithm in $K[x]$, Maza and Rioboo multiply $r_i$ by a
quasi-inverse of $u$ before pseudo-division by $r_{i-1}$.
Suppose ${\rm den}(r_i)=1$ and let $v$ be a quasi-inverse of $u={\rm lc}(r_i)$.
Then ${\rm den}(v r_i)=1$ and ${\rm lc}(v r_i) \in \ZZ$ thus
quantity $\mu$ in the pseudo-division will be an integer.
We obtain the following $\ZZ$-fraction-free algorithm for computing an
associate of the monic gcd $g$ of $f_1, f_2 \in K[x].$

\bigskip
\noindent
\hspace*{5mm} 1  Set $r_1 = \hat f_1$, $r_2 = \hat f_2$. \\
\hspace*{5mm} 2  Compute $v$ s.t. $v u = r$ for $r \in \ZZ$ where $u = {\rm lc}(r_1)$ and set $r_1 = v r_1.$ \\
\hspace*{5mm} 3  Set $i=2.$ \\
\hspace*{5mm} 4  Compute $v$ s.t. $v u = r$ for $r \in \ZZ$ where $u = {\rm lc}(r_i)$ and set $r_i = v r_i.$ \\
\hspace*{5mm} 5  \hspace*{5mm} Let $\bar r$ be the pseudo-remainder of $r_{i-1}$ divided $r_i$ mod $T$. \\
\hspace*{5mm} 6  \hspace*{5mm} If $\bar r=0$ then output $r_i$. \\
\hspace*{5mm} 7  \hspace*{5mm} Set $i = i+1$ and $r_i = \bar r$ and go to step 4.

\bigskip
\noindent Although this algorithm is $\ZZ$-fraction-free the size of
the integer coefficients blows up exponentially.  This is caused by
multiplication by the integer $\mu$ in pseudo-division and also by
multiplication by $r$ when multiplying by the quasi-inverse $v.$
This blowup can be reduced either by dividing out by known integer
factors, which is the approach that Maza and Riboo take in
\cite{Maza} in modifying the subresultant GCD algorithm, or by
making $r_i$ and $\bar r$ primitive, that is, dividing out by the
gcd of their integer coefficients. Which approach is better depends
on the relative cost of computing gcds verses multiplication and
division in the base coefficient domain which in our case is $\ZZ$.
We recall the notion of integer primitive part and integer content
for $K[x]$.

\begin{definition}
Let $f \in K[x]$ with ${\rm den}(f) = 1$.
The integer content of $f,$ denoted ${\rm ic(f)}$ is the gcd of the
integer coefficients of $f$ when $f$ is viewed as a polynomial in $\ZZ[z_1,..,z_n][x]$.
The $\ZZ$-primitive part of $f,$ denoted ${\rm pp}(f)$ is $f/{\rm ic}(f).$
Thus we have $f = {\rm pp}(f) ~ {\rm ic}(f)$ and ${\rm pp}(f) = \hat f$.
\end{definition}
After computing $r_i = v r_i$ in step 4 we set $r_i = {\rm pp}(r_i)$ and
also after computing $\bar r,$ in step 5 we set $\bar r = {\rm pp}(\bar r)$.
The resulting GCD algorithm that we obtain is a {\em primitive $\ZZ$-fraction-free}
algorithm.

It remains to describe how we compute a quasi-inverse of $u \in K.$
One way to do this would be to compute $u^{-1}$ using the extended Euclidean algorithm
applied to $m_n$ and $u$ in $K_{n-1}[z_n]$ and then clear fractions.
In the same way we have just described how to modify the monic Euclidean algorithm
for computing a gcd in $K[x]$ where $K = K_{n-1}[z_n]/\langle m_n \rangle,$
to be $\ZZ$-fraction-free, Maza and Rioboo modify the extended monic Euclidean
algorithm in $K_{n-1}[z_n]$ to be fraction-free by using pseudo-division
and multiplication by quasi-inverses in $K_{n-1}$.
Again, an eponential blow up occurs which can be reduced by dividing out by
known integer factors or it can be minimized by dividing out by integer contents.
To fix the details of this algorithm we present our Maple code for computing
the quasi-inverse of $u \in K$ and integer $r$ using our Maple data structure
from the previous section.

{ \small
\begin{verbatim}
  quasiInverse := proc(x) local Q,K,P,m,u,r0,r1,t0,t1,i,c,den,g,pr,mu,pq;
  # Input  x in K = K_{n-1}[z]/<m(z)>
  # Output v in K and r in Z^+ s.t.  v x = r and r = den(1/x)
    Q := [0,[],[]];  # field of rational numbers
    K := getring(x);
    if K=Q then u := rpoly(x); RETURN( rpoly(denom(u),Q), numer(u) ) fi;
    m := getalgext(K);    # m is a polynomial in z
    u := liftrpoly(x);    # u is a polynomial in z
    u := ipprpoly(u,'c'); # x = c u and u for u primitive over Z
    P := getring(m);      # P = K[i-1][z]
    r0,r1,t0,t1 := m,u,rpoly(0,P),rpoly(1,P);
    while degrpoly(r1) > 0 do
        (i,den) := quasiInverse(lcrpoly(r1));
        (r1,t1) := mulrpoly(i,r1),mulrpoly(i,t1);
        g := igcd(icontrpoly(r1), icontrpoly(t1));
        (r1,t1) := iquorpoly(r1,g),iquorpoly(t1,g);
        pr := ippremrpoly(r0,r1,'mu','pq');
        if iszerorpoly(pr) then ERROR( "inverse does not exist", [r1,P] ) fi;
        r0,r1,t0,t1 := r1,pr,t1,subrpoly(mulrpoly(mu,t0),mulrpoly(pq,t1));
        g := igcd(icontrpoly(r1), icontrpoly(t1));
        (r1,t1) := iquorpoly(r1,g),iquorpoly(t1,g);
    end while;
    (v,r) := quasiInverse(lcrpoly(r1),args[2..nargs]);
    t1 := mulrpoly(v,t1); g := igcd(icontrpoly(t1),r);
    t1 := scarpoly(denom(c),iquorpoly(t1,g));
    RETURN( subsop(1=K,t1), numer(c)*r/g );
  end;
\end{verbatim}
}
We remark that at the start of the loop we have $s m + t_1 u = r_1$ for
some $r_1$ in $K_{i-1}[z]$ (and some $s \in K_{i-1}[z]$ which is not computed).
Thus when the loop exits we have $t_1 u \equiv r_1 \bmod m$ for a constant
polynomial $r_1 \in K_{i-1}[z]$.
We multiply $t_1$ by $v$ the quasi-inverse of $r_1$ so that we
have $t_1 x \equiv r \bmod m$ for some $r \in \ZZ$.
But multplication by $v$ introduces an integer multiplier and since this
algorithm algorithm is recursive it is critical that we clear it here.
Thus we compute $g$ the gcd of $r$ and the coefficients of $t_1$ and divide
through by $g$.

We can improve the performance of this algorithm further by
modifying pseudo-division as follows; instead of multiplying $f_1$ by $\mu$
and then performing a normal long division, we modify the division algorithm to
multiply the current pseudo remainder $r_{i-1}$ by the smallest integer s.t. the
leading coefficient of the divisor $r_i$ will divide the leading coefficient of $r_i$ exactly.
We call this $\ZZ-$primitive pseudo-division.
This is what the subroutine {\tt ippremrpoly} does.
This improvement gives us typically another 30\% improvement in quasi-inverse computation.

Finally, what if $K$ is not a field?
Suppose we call the algorithm with $u \in K.$
If the algorithm returns normally, it outputs $v \in K$ and $r \in \ZZ$
such that $v u = r$.  Then $u$ is invertible for $u^{-1} = v/r$.
Suppose an error occurs and the algorithm outputs $g,P$.
Then $g \in P = K_{i-1}[z_i]$ for some $1 \le i < n$ is a non-trivial
factor of $m_i \in P$ and thus we have encountered a zero divisor
$w \in K_i.$

Of the two, Maza and Rioboo's algorithm and our primitive
$\ZZ$-fraction-free algorithm which is derived from Maza and
Rioboo's algorithm, ours appears to be much faster.  In
\cite{MonvHoeij}, we showed that there is a cubic growth in the size
of the integers in Maza and Rioboo's algorithm whereas the growth in
the primitive $\ZZ$-fraction-free algorithm is linear.

%% file: timings.tex
%
%
%
%

\subsection{Timing Results}

In this section we compare the Magma and Maple
implementations of our modular GCD algorithm with the default
Maple and Magma system GCD implementations for a sequence of
univariate gcd problems over a number field $L$ of degree 24.
The number field $L = \QQ(\alpha,\beta)$ used in our test
problems is defined by $m_\alpha(z) = z^8-40 z^6+352 z^4-960 z^2+576$ and
$m_\beta(z) = z^3-11 z-13$.  The gcd problems are constructed as follows.
Let
\[ g = x^2+123 \beta x+\alpha x/13+531 \alpha^3-199,\]
\vspace*{-5mm}
\[ a = x^2+\alpha x/12+123 \beta-25 \alpha^3+251, ~~ {\rm and} \]
\vspace*{-5mm}
\[ b = x^2+\beta/21+123 \alpha x+17 \alpha^3-173. \]
For $k = 0, 1, 2, ..., n$ the input polynomials $f_1$ and $f_2$
are defined as follows: $f_1 = g^k a^{n-k}$ and $f_2 = g^k b^{n-k}.$

Thus we consider a sequence of gcd problems over $L$ where the degree of
the input polynomials is fixed at $2 n$ and the ${\rm gcd}(f_1,f_2) = g^k$,
is a polynomial of degree $2 k$.
The reason for this choice of gcd problems, where
the degree of the gcd is increasing relative to the
degree of the inputs, is that it includes a range of types of gcd problem
that occur in practice.  In comparison with the Euclidean algorithm,
we expect our modular GCD algorithm to perform best for small $k$ and
worst for large $k$.

The following comparison is made between Maple 9 and Magma 2.10 on
an AMD Opteron running at 2.0 GHz for $n=10$, that is, the degree of
the input polynomials $f_1$ and $f_2$ is 20. All timings are in CPU
seconds. The timings in columns 1 and 4 are for our Maple and Magma
implementations of our modular Gcd algorithm where the number of
primes required for reconstruction is indicated in parens. 
The Maple timings in column 2 are for the monic Euclidean algorithm.
The Maple timings in column 3 are for the primitive fraction free GCD algorithm.
The Magma timings in columns 5 and 6 are for the monic Euclidean
algorithm over $L$ where the elements of $L$ are created using
Magma's NumberField constructor (column 5) and Magma's quotient
field constructor (column 6).

\begin{table}[htb]
\begin{center}
\begin{tabular}{r | l r r | r r r |}
   & Maple & Rel 9 &     & Magma & 2.10  & \\
$k$ &  1       &   2         &  3     &    4       &   5   &   6   \\ \hline
 0  & 0.27 (1) &   NA        & NA     &  0.06 (1)  & 200.4 &    NA \\
 1  & 1.3  (3) &   NA        & NA     &  0.09 (2)  & 151.6 &    NA \\
 2  & 1.5  (4) &   NA        & NA     &  0.12 (3)  & 106.1 &    NA \\
 3  & 2.4  (6) & 367.3       & NA     &  0.16 (4)  &  66.1 &    NA \\
 4  & 3.1  (9) & 193.7       & NA     &  0.20 (5)  &  37.8 &    NA \\
 5  & 3.4 (11) &  90.0       & NA     &  0.21 (6)  &  18.3 & 808.5 \\
 6  & 3.4 (13) &  37.7       & NA     &  0.20 (7)  &   7.3 & 282.4 \\
 7  & 3.1 (15) &  13.0       & 176.2  &  0.19 (8)  &   2.1 &  73.7 \\
 8  & 2.4 (17) &   3.5       &  39.5  &  0.19 (10) &   0.4 &  12.4 \\
 9  & 1.6 (19) &   0.8       &   2.0  &  0.14 (11) &   0.1 &   1.1 \\
10  & 1.0 (22) &   0.1       &   0.0  &  0.11 (12) &   0.0 &   0.0 \\ \hline
\end{tabular}
\caption{NA means not attempted}
\end{center}
\end{table}

\subsubsection*{Remarks}
\begin{enumerate}
\item The number of primes (indicated in parens) for
  the modular algorithm is more in Maple than in Magma.
  This is because Maple 9 uses 15.5 bit primes for portability \cite{Disco}
  and Magma 2.10 uses 30 bit primes \cite{Steel}.

\item Even allowing for the fact that Magma uses fewer primes than
  Maple, the Magma implementation is considerably faster.
  The Maple implementation is using compiled code for arithmetic
  in $\ZZ$, $\QQ$ and $\ZZ_p[z]$ but not for rational reconstruction,
  nor arithmetic in $\QQ[x]$ and $\ZZ_p[z]/\langle m(z) \rangle[x]$
  wheres Magma does.  Thus less time is spent in the Magma interpreter.

\item The times for both implementations increase to a
  maximum at $k=6$ then decrease even though the number of primes
  increases linearly.  The reason is that the cost of the modular gcds,
  which is $O( (n/2+k/2+1)(n+1-k) )$ coefficient operations,
  is decreasing {\em quadratically} to $O(n+1)$ as $k$ increases to $n$,
  and the cost of the trial divisions, which is $O( (k+1)(n+1-k) )$
  coefficient operations in $L$, is also decreasing after $k=n/2$
  {\em quadratically} to $O(n+1)$.

\item The reason for the huge difference in times between columns
  5 and 6 is because of the different representation of field elements
  being used and the different algorithm for inverting field elements.
  In column 5 we used the {\tt NumberField} constructor to build $L$
  which represents field elements as as polynomials over $\ZZ$
  {\em with denominators factored out}.  In column 6 we have used
  the quotient ring constructor to build $L$ which doesn't.
  Thus {\tt NumberField} avoids arithmetic with fractions.
  The second reason is that {\tt NumberField} uses a modular
  algorithm to compute inverses in $L$ which is where, by experiment,
  most of the time is spent on this data.

\item The data clearly shows the superiority of the modular GCD algorithm.
  And yet the non-modular timings are still impressive.  This is partly because
  Maple 9 and Magma 2.10 both have asymptotically fast integer arithmetic.
  However, the data also shows that the Euclidean algorithm is faster
  than the modular GCD algorithm when $\deg (g)$ is large.
  The efficiency of the modular GCD algorithm can be improved when $g$ is large
  if we reconstruct also $f_2/g,$ the smaller cofactor.
  We will show timings for this next.
\end{enumerate}

The second set of data below is for the same gcd problem set but
with $n=15$ instead of $n=10$. The four sets of timings, all in CPU
seconds, in columns 0, 1, 3, and 5 are for the primitive
$\ZZ$-fraction-free algorithm and for three versions of our Maple
implementation of the modular GCD algorithm. In column 1 we are
using Wang's rational reconstruction algorithm (see \cite{Wang81}).
In column 3 we are instead using Monagan's maximal quotient rational
reconstruction algorithm (MQRR) from \cite{MQRR}. In column 5 we
also reconstruct the smaller cofactor, stopping when rational
recocnstruction succeeds on the cofactor or the gcd. In columns 2, 4
and 6, the first number in parens is the number of (good) primes
that the modular GCD algorithm and second number indicates the time
spent in trial division.

\begin{table}[htb]
\begin{center}
\begin{tabular}{r r | r r | r r | r r}
$k$& 0    &    1  &  2         &    3   &  4         &    5   & 6    \\ \hline
0& -     & 0.659 &(1, 0\%)  &  0.669 &(1, 0\%)   & 0.660 & (1, 0\%)  \\
1& -     & 1.631 &(2, 17\%) &  1.621 &(2, 17\%)  & 1.599 &(2, 16\%)  \\
2& -     & 2.679 &(3, 25\%) &  2.681 &(3, 25\%)  & 2.710 &(3, 26\%)  \\
3& -     & 4.529 &(5, 29\%) &  3.860 &(4, 33\%)  & 3.860 &(4, 34\%)  \\
4& -     & 6.859 &(8, 27\%) &  5.600 &(6, 33\%)  & 5.689 &(6, 34\%)  \\
5& -     & 7.910 &(10, 24\%)&  6.590 &(7, 37\%)  & 7.500 &(7, 33\%)  \\
6& -     & 10.32 &(12, 28\%)&  7.350 &(8, 39\%)  & 8.449 &(8, 33\%)  \\
7& -     & 11.33 &(14, 28\%)&  7.820 &(9, 38\%)  & 9.140 &(9, 33\%)  \\
8& -     & 11.68 &(16, 27\%)&  8.000 &(10, 37\%) & 9.350 &(10, 32\%)  \\
9&  268. & 11.71 &(18, 25\%)&  7.809 &(11, 36\%) & 8.119 &(9*, 34\%)  \\
10& 126. & 11.67 &(21, 23\%)&  7.430 &(12, 33\%) & 6.659 &(8*, 37\%)  \\
11& 53.2 & 10.19 &(22, 21\%)&  6.559 &(13, 30\%) & 4.530 &(6*, 43\%)  \\
12& 19.1.& 8.840 &(25, 17\%)&  5.539 &(14, 25\%) & 3.030 &(5*, 44\%)  \\
13& 5.49.& 9.170 &(27, 8\%) &  4.170 &(15, 17\%) & 1.470 &(3*, 52\%)  \\
14& 1.21 & 4.120 &(29, 6\%) &  3.310 &(17, 8\%)  & 0.580 &(2*, 52\%)  \\
15& 0.14 & 2.030 &(31, 0\%) &  2.259 &(18, 0\%)  & 0.149 &(2*, 34\%)  \\ \hline
\end{tabular}
\caption{(*) means cofactor reconstructed and (-) not attempted.}
\end{center}
\end{table}


%% file: conclusion.tex
\section{Conclusion and Remaining Problems}
Let $L$ be a number field of degree $D$ presented with $\ls$ field extensions.
Let $f_1, f_2 \in L[x]$ and let $g$ be the monic gcd of $f_1$ and $f_2$.
We have presented a modular GCD algorithm which computes $g$ without
converting to a single field extension and without computing discriminants (Thereom 1).
Our goal was to design an algorithm with a complexity
that is as good as classical polynomial multiplication and
division in $L[x].$ Recall that $H(g)$ denotes the magnitude of the
largest integer appearing in the rational coefficients of $g \in L[x].$
Let $m = \log H(g)$ and $M = \max( \log H(f_1), \log H(f_2) ).$
Our algorithm incrementally reconstructs $g$ from its image modulo $k$ machine
primes such that $k$ is proportional to $\log H(g).$
It uses rational reconstruction.
The reason for using an incremental approach with trial division
rather than using a bound is that there are no good bounds for $H(g),$
in particular, when $H(g)$ is much smaller than $\min( H(f_1), H(f_2) )$.

Our implementations of the algorithm in Maple and Magma demonstrate its
effectiveness compared with non-modular algorithms.
Both implementations use a recursive dense representation
for field elements and for polynomial variables to eliminate data
structure overhead in the algorithm which otherwise may
ruin a modular implementation.

Our Maple implementation was installed in Maple 10 in 2005
as part of the {\tt Algebraic} package by J\"urgen Gerhard of Maplesoft.
It may be accessed using the Maple command
{\tt with(Algebraic:-RecursiveDensePolynomials);}
We made one further optimization that we found useful.
Many applications in practice involve algebraic numbers which are
simple square roots and cube roots such as $i=\sqrt{-1}$ and $\sqrt 2$.
In such cases it is advantageous to pick primes for which the 
the minimal polynomial $m_1(z)$ for $\alpha_1$ splits into distinct 
linear factors modulo $p$.  For if $m_1(z) = \Pi_{j=1}^d (z-\beta_j)$ in $\ZZ_p[z]$
then we may compute the gcd of $f_1(z=\beta_j)$ and $f_2(z=\beta_j)$
for each $j$ and interpolate $z$.  
This embeds $\alpha_1$ in $\ZZ_p$ and eliminates a field extension.
In practice, it eliminates computations with polynomials in $z$ of
low degree which have a relatively high data structure overhead.
In our software we do this if $\deg m_1(z) \le 4$ where
there is a reasonable probability of finding primes that split $m_1(z)$.


Write $L = \QQ(\alpha_1,\ldots,\alpha_\ls)$ and $D = d_1 \cdots d_\ls$
where $d_i$ is the degree of the minimal polynomial of $\alpha_i$.
If the degree $D$ of $L$ over $\QQ$ is high then the use of fast
multiplication techniques can speed up the arithmetic in $L$ mod $p$.
In particular, if $\ls=1$ we can multiply and divide
in $\ZZ_p[z]$ in $O( \tilde D = D \log D \log \log D).$
To multiply polynomials in $L_p[x]$ where $L_p = \ZZ_p[z]/\langle m(z) \rangle$
rapidly, first multiply them as bivariate polynomials in $\ZZ_p[z,y]$ then reduce
the coefficients modulo $m(z)$ using asymptotic fast division.
To multiply the bivariate polynomials rapidly, first convert them, in linear
time to univariate polynomials using the subtitution $y \rightarrow z^D$.
This large multiplication in $L_p[z]$ is in $O( N D ( \log (ND) \log \log (ND) ) )$.
Now a fast multiplication in $L_p[x]$ enables a fast GCD computation
in $L_p[x]$ in $O( N D \log^2 (ND) \log \log (ND) ) ) = O(\tilde{N} \tilde{D})$.
Thus for $\ls=1,$ we have sketched out an asymptotically modular GCD
algorithm which runs in $O( \tilde{M} N D + m \tilde{N} \tilde{D} )$ time
with high probability.

If $\ls>1,$ asymptotically fast multiplication and division in $L_p$ will
be less effective than if $\ls=1$.  This suggests that we first convert
to a single field extension.  Using a primitive element in characteristic 0
should be avoided because it can cause coefficient growth.
But in characteristic $p$ this is not a concern.
Let $\gamma = c_1 \alpha_1 + c_2 \alpha_2 + ... + c_\ls \alpha_\ls$
be a primitive element.  Using linear algebra one can compute the
minimal polynomial $m(z) \in \ZZ_p[z]$ for $\gamma,$ and also the
representation for all $D$ power products $\alpha_1^{e_1} \times ... \times \alpha_\ls^{e_\ls}$
in $\ZZ_p[z]$ in O($D^3$) arithmetic operations in $\ZZ_p$
and then make the substitutions for the power products in $f_1$
and $f_2$ in O($N D^2$) arithmetic operations in $\ZZ_p$.
If $N,$ the degree of $f_1$ and $f_2$ is high enough, the
time saved by the fast multiplication techniques in the Euclidean
algorithm mod $p$ will be larger than the cost of the
conversion to a single extension mod $p$.
However, if we want to do this then we really need also
to think about how to convert to a single field extension faster
than $O( D^3 + N D^2 ).$  We do not know how to do this.